\documentclass[a4paper, amsfonts, amssymb, amsmath, reprint, showkeys, nofootinbib,superscriptaddress,twocolumn ]{revtex4-2}
\usepackage{float}
\usepackage{graphicx}
\usepackage{longtable}
\usepackage[left=15mm,right=13mm,top=23mm,columnsep=15pt]{geometry} 
\usepackage[caption=false]{subfig}
\usepackage{hyperref} 
\begin{document}
\title{Microscopic study of strange hadron productions at the LHC with$\sqrt{s_{NN}}$=2.76 TeV}

\author{Purabi Ghosh}
    \affiliation{Department of Applied Physics and Ballistics, Fakir Mohan University, Balasore-756019, India}
\author{Jajati K. Nayak}
    \email[Correspondence email address: ]{jajati-quark@vecc.gov.in}
    \affiliation{Variable Energy Cyclotron Centre, 1/AF, Bidhan Nagar , Kolkata, India}
\author{Sushant K. Singh}
    \affiliation{Variable Energy Cyclotron Centre, 1/AF, Bidhan Nagar , Kolkata, India}
    \affiliation{HBNI, Training School Complex, Anushakti Nagar, Mumbai 400085, India}
\author{Santosh K. Agarwalla}
    \affiliation{Department of Applied Physics and Ballistics, Fakir Mohan University, Balasore-756019, India}

\date{\today} 

\begin{abstract}
Ratio of the yield of strange hadrons to pions is considered as an important observable in studying the properties of the system produced in relativistic heavy ion collisions. Production of strange hadrons $K, \bar{K}, \Lambda, \Sigma, \Xi$ and $\Omega$ have been evaluated microscopically using rate equations by considering their hadronic interaction cross sections in an expanding medium. The yields obtained from rate equations are normalized with thermal pions and compared with the measurements from Pb-Pb collisions at various multiplicities at LHC energy. The calculation has been done for various initial and freeze out conditions. At 2760 GeV, LHC energy, $\Lambda, \Xi, \Omega$ freeze out close to $T_C$ and $K^0_s$ freezes out little later. But there is a subtle difference in freeze out temperatures of different species which may be distinguishable at lower colliding energies. 
\end{abstract}


\maketitle 
\section{Introduction}
The study of strange hadrons plays a key role in extracting the properties of the medium produced in relativistic heavy ion collisions. The experiments performed at several colliding energies in several accelerator facilities like Alternating Gradient Synchrotron(AGS), Relativistic Heavy Ion Collider (RHIC), Super Proton Synchrotron(SPS) and Large Hadron Collider(LHC) provide ample of strange hadron data which help understand the QCD phase diagram. Recently, ALICE collaboration has measured the yield of strange hadrons $K, \Lambda, \Xi$ and $\Omega$ in p-p collisions at $\sqrt{s_{NN}}$=7 TeV, p-Pb collisions at $\sqrt{s_{NN}}$=5.02 TeV \cite{alicenature17,gyula_alice} and Pb-Pb collisions at $\sqrt{s_{NN}}$=2.76 TeV  \cite{multistrange_alice_plb14} in various centralities. The normalised yields of strange hadrons, ($H_s+\bar{H_s}$)/($\pi^++\pi^-$) are measured at various charge particle multiplicities and presented as an observable by ALICE collaboration \cite{alicenature17}.
The measured data in \cite{alicenature17}  are $(2K^0_s)/(\pi^++\pi^-)$, $(\Lambda+\bar{\Lambda})/(\pi^++\pi^-)$, $(\Xi^-+\bar{\Xi^+})/(\pi^++\pi^-)$ and $(\Omega^-+\bar{\Omega^+})/(\pi^++\pi^-)$. We call the ratio $(H_s+\bar{H_s})/(\pi^++\pi^-)$ as the '{\it yield-ratio}' throughout the article for the sake of convenience. The {\it yield-ratios} show a smooth increasing pattern with multiplicity and then a saturation for all strange hadrons but with a little deviation for $\Xi/\pi$ at lowest measured multiplicity and also at highest multiplicity.

These measurements are extremely important, as  a smooth pattern of {\it yield-ratio} with charged multiplicity, from various colliding systems (p-p, p-Pb, Pb-Pb) at different colliding energies, would answer the question of similarity of systems with similar multiplicities produced in these collisions and a deviation may hint for new physics. The {\it yield-ratio} for $\Lambda$, $K$, $\Omega$ show a smooth increasing pattern, but as mentioned above, $\Xi$ shows a deviation. It is also observed that slopes of {\it yield-ratio} for multi strange hadrons are more compared to single strange hadrons. This may signify the enhancement of multi-strange productions compared to single-strange ones. To analyze the phenomenon, understanding of the microscopic mechanism for the production of all strange hadrons is necessary, which is the focus of this calculation.

The study of strange hadrons was believed to be important because of their enhanced productions in heavy ion (A-A) collisions over proton-proton(p-p) collisions and was proposed long before as a good signature of quark gluon plasma (QGP) ~\cite{rafelski82,GG99} formation. Widely discussed horn like structure in the measurements of $K^+/{\pi^+}$ ~\cite{na49alt,star,brahms,na49afa} ratio with colliding energies ignited many theoretical models in the last two decades. The multi strange baryons $\Xi(uss,dss)$ and $\Omega(sss)$ also show enhancement like $K^+$ in  Pb-Pb collisions at $\sqrt{s_{NN}}$=2.76 TeV \cite{multistrange_alice_plb14} over p-p collisions. The observations of $\Xi$ \& $\Omega$ yield at  $\sqrt{s_{NN}}$=200 GeV, Au+Au collisions\cite{multistrange_star_prc08} also supported the argument of strangeness enhancement in A-A collisions. Similar observations have also been made at SPS energy by WA97 collaborations while measuring $\Xi$ and $\Omega$ from Pb-Pb and p-Pb collisions at CERN \cite{andersen99}. Enhancement in case of $\Xi, \Omega$ is a factor of 3 in Pb-Pb over p-p collisions at 158 A GeV\cite{andersen99}. $\Xi/\pi$ and $\Omega/\pi$ in Pb-Pb collisions at $\sqrt{s_{NN}}$=2.76 TeV  are  1.6 and ~3.3 times more compared to  p-p collisions at $\sqrt{s_{NN}}$=7 TeV at LHC\cite{multistrange_alice_plb14}. In the mean time, the availability of most recent data from p-Pb collision at $\sqrt{s_{NN}}$=5.02 TeV \cite{cascadealice15} makes it more interesting as it would help in providing a systematic study from p-p to p-A to A-A collisions. 

Strangeness productions in QGP and hadronic phases have been studied by various models by several authors~\cite{rafelski82,kapusta86,cugnonprc90,LR08,RL99,BT07,sgupta10,CORS,MG04,andronic06,jknacta06,jknprc10,tawfik09}. The enhanced production of kaons and anti-kaons in the experiments at various colliding energy ranges such as SIS( up to 1 GeV)\cite{sis94}, AGS (up to 10 A GeV) \cite{ags91} and SPS energies (~11-158 A GeV)\cite{sps98} are explained by hadronic scatterings in some of the above works \cite{jknprc10,BT07} and also using AMPT \cite{ampt_ko_96}. However the multi strange baryon productions 
have not been explained satisfactorily. 

Statistical Hadronisation Model also evaluated the integrated yield at those energies assuming common chemical freeze out temperature for all species\cite{sgupta10,andronic06,andronicplb09} including RHIC and LHC energies. However, the model could not explain the ratios of multi strange hadrons for 0-20\% centrality of Pb-Pb collision at 2.76 TeV LHC energy  while fitting with $p/\pi$ ratio. Similarly productions of kaons and anti-kaons at higher colliding energies such as at RHIC and LHC (also at higher SPS energies) have been explained using models with strange quark evolution assuming a QGP phase\cite{BT07,jknprc10}. But the multi strange productions are not explained there. 

Using minimal statistical hadronization model, the authors in \cite{kolomeitsev12} tried to explain the momentum spectra of hyperons $\Lambda, \Xi$ measured by HADES collaboration from the collision of Ar at 1.76A GeV on fixed target KCl without considering the microscopic productions. Same authors\cite{kolomeitsev12} also explained the kaon productions ($K^+/\pi^+$ vs $\sqrt{s_{NN}}$)from NICA experiment (NICA white paper) but failed to reproduce the ratio $\Xi^-/\Lambda$ and $\Omega^-/\Xi^-$ simultaneously\cite{kolomeitsev15}.  However they got a similar trend of $\Xi^-/\Lambda$ and $\Omega^-/\Xi^-$ with $\sqrt{s_{NN}}$ although the calculation under predicts the data. The authors in \cite{steinheimerjpg16} provided a possible explanation of subthreshold production of $\Xi^-$ by considering new decay channels of massive baryon resonances. 

An attempt has also been made in \cite{bassplb99} to explain the multi strange productions at SPS energy using Ultra Relativistic Quantum Molecular Dynamics(UrQMD), but data were not reproduced. It is observed nowadays that people are using sequential freeze out scenario to explain the strange hadron yields. Rene Bellewied in one such attempt while discussing sequential freeze out of strange hadrons argued in favor of flavor dependent freeze out by comparing latest lattice computation and data of net-kaon, net-charge and net-proton fluctuation\cite{bellewiedepj18}.

In this paper, the microscopic productions of $K,\bar{K}, \Lambda, \Sigma, \Xi$ \& $\Omega$ have been discussed with their interactions in the hot-dense system along with their evolutions considering Bjorken expansion and using rate equation.  We focus our calculation to analyse the {\it yield-ratio} data for all strange hadrons at different multiplicities from Pb-Pb collision at $\sqrt{s_{NN}}$= 2.76 TeV. Since the hadron gas produced in heavy ion collision is supposedly a dilute gas, hence consideration of rate equation or transport calculation is very much relevant. But instead of full (3+1) dimensional expansion, we have considered relatively easier Bjorken expansion as we are interested in the ratio of the numbers. The final numbers unlike momentum $p_T$ spectra or flow observable would not change much if we employ (3+1)dimensional expansion. However the freeze out parameters may change quantitatively although qualitative change is not expected. To compare, a calculation with 3-d Hubble expansion which is relevant in hadronic phase is under progress. Full (3+1)-d hydrodynamical treatment is kept for the future work.

We divide the manuscript as follows. The cross-sections of productions for Kaon, Lambda, Sigma, Cascade and Omega in a hadronic medium are discussed  in section~\ref{sec:crosssection}.  The formalism for rate of production is described in Section~\ref{sec:rate}. The rate equations for single- and multi-strange hadrons are discussed considering a Bjorken expansion in section~ \ref{sec:rateequation}. The evolution equations for temperature and baryon chemical potential($\mu$) are also described here. Then the results are presented in the section-V and finally, section~\ref{sec:summary} is devoted to summary and conclusion.
\section{\label{sec:crosssection} Production and interaction of strange hadrons in hadronic medium}
In case of relativistic heavy ion collisions, the observed hadrons might be produced either due to hadronisation of quarks when initial quark gluon state is produced or because of the nucleonic interactions of the colliding nuclei. The yield in the later case would be low. In both the cases, produced hadrons undergo further scatterings inside the medium till they decouple and free stream towards the detector. The dynamics of hadrons determines the properties of the system and hence the final yield. In this study, the aim is to understand the dynamics of strange degrees of freedom. While considering the production and interaction of strange hadrons we assume the non-strange hadrons to provide thermal background. The time evolution of the hadronic system is studied with rate equation or momentum integrated Boltzmann equation along with Bjorken expansion of the system.

Various interactions involving strange and non-strange hadrons that produces $K,\Lambda, \Sigma, \Xi$ and $\Omega$ are discussed below.

\subsection{Interaction channels and strange hadron cross sections}
The production of strange mesons $K, \bar{K}$ and baryons $\Lambda, \Sigma, \Xi, \Omega$ are studied with the following hadronic interactions. They can be categorised as meson-meson(MM), meson-baryon(MB) and baryon-baryon(BB) interactions based on the hadrons in the initial channel. $B$ and $Y$ represents non-strange baryons and hyperons respectively. The reactions are, 
$\pi \pi \rightarrow K \bar{K}$,
$\pi N \rightarrow \Sigma K$, 
$ \bar{p} p \rightarrow \Lambda \bar{\Lambda}$,
$\pi \rho \rightarrow K \bar{K}$, 
$\bar{K} N \rightarrow \Lambda \pi$,  
$ \bar{p} p \rightarrow \Sigma^- \bar{\Sigma^+}$, 
$\rho \rho \rightarrow K \bar{K}$, 
$\bar{K} N \rightarrow \Sigma \pi$,   
$\pi N \rightarrow \Lambda K$, 
$\rho N \rightarrow \Lambda K$, 
$ \bar{p} p \rightarrow K^- \bar{K^+}$,
$ \bar{K}N \rightarrow K \Xi$, 
$ \bar{K}\Lambda \rightarrow \pi \Xi $,
$ \bar{K}\Sigma \rightarrow \pi \Xi $, 
$ \Lambda\Lambda \rightarrow N \Xi$,
$ \Lambda\Sigma \rightarrow N \Xi $,
$ \Sigma\Sigma \rightarrow N \Xi$,
$ \bar{K}\Sigma \rightarrow \pi \Xi $, 
$\Lambda \bar{K}\rightarrow \Omega^{-} K^0$,
$\Sigma^{0} \bar{K}\rightarrow \Omega^{-} K^0$,
$\bar{p} p\rightarrow \Omega^{-} \Omega$, 
$ p\bar{p} \rightarrow \Xi \bar{\Xi}$.
where, $N$ represents nucleons(proton or neutron). The production cross section for hadrons with single strangeness is described in ~\cite{jknprc10,Brown1,amslar08,patrignani16,liprc12,kaidalov94,cugnonnpa84,linpa97}. Many of them are verified with experimental observations. The cross sections for inverse reactions are also taken into account using principle of detailed balance as in ~\cite{cugnon84}. There are some $2\rightarrow 3$ channels involving baryons in the initial channels such as $BB\rightarrow BYK$ which might be relevant for strange production at low colliding energies that is when baryon density in the system is high. However, we neglect contributions from such processes due to phase space factor. 
\subsection{Production cross sections for single-strange hadrons}
Among the strange hadrons carrying single strange quantum number, Kaons ($K, \bar{K}$) are the lightest one. For Kaon production, the isospin averaged cross section ($ab\rightarrow cd$) from MM interactions ($\pi\pi\rightarrow K\bar{K}$, $\rho\rho\rightarrow K\bar{K}$, $\pi\rho\rightarrow K\bar{K^*}$ and $\pi\rho\rightarrow {K^*}\bar{K}$) is given by,
\begin{equation}
{\bar{\sigma}}_{ab\rightarrow cd}(s)=\frac{1}{32 \pi} \frac{P'_{cd}}{sP_{ab}} \int^1_{-1}dx M(s,x)
\end{equation}
where, $s=(p_a+p_b)^2$ with $p_a, p_b$ being the four momenta of incoming particles $a$ and $b$; $P_{ab}$ and $P'_{cd}$ are three momenta of incoming mesons and outgoing kaons in the centre-of-mass frame, $x= {\text cos}(P_{ab},P'_{cd})$. $M$ is invariant amplitude and calculated from following interaction Langrangian densities~\cite{Brown1}, 
${\mathcal L}_{K^*K\pi}=g_{K^*K\pi}K^{*\mu}\tau[K(\partial_{\mu}\pi)-(\partial_{\mu}K)\pi]$ and ${\mathcal L}_{\rho KK}=g_{\rho KK}[K\tau(\partial_{\mu}K)-(\partial^{\mu}K)\tau K]\rho^{\mu}$. 
Similar to MM interactions, MB interactions ($MB\rightarrow YK$) also produce Kaons but strange baryons such as $\Lambda$ and $\Sigma$ are also produced along with. The dominant contributions in this category come from $\pi N \rightarrow \Lambda K$, $\pi N \rightarrow \Sigma K$, $\rho N \rightarrow \Lambda K$, $\rho N \rightarrow \Sigma K$. The production cross sections are evaluated and parametrised in \cite{Brown1,cugnonnpa84}. We have calculated the cross section using the following expression by considering $N^*_1(1650), N^*_2(1710), N^*_3(1720)$ \cite{amslar08} and $N^*_5(1875), N^*_6(1900)$ \cite{patrignani16} as intermediate resonant states.

\footnotesize
\begin{align}
\bar{\sigma}_{MB\rightarrow YK} &= \nonumber \\ 
&\sum_i\frac{(2J_i +1)}{(2S_1+1)(2S_2+1)} \frac{4\pi}{k_i^2}\frac{\Gamma_i^2/4} {(s^{\frac{1}{2}}-m_i)^2+\Gamma_i^2/4} B_i^{\text{in}} B_i^{\text{out}}
\end{align}\normalsize
The sum is over resonances with mass($m_i$), spin ($J_i$) and decay width($\Gamma_i$). $(2S_1+1)$ and $(2S_2+1)$ are the polarisation states of the meson (M) and baryon(B) in the incoming channels. $B_i$ represents the branching ratio. Other required parameters are taken from the particle data book \cite{amslar08,patrignani16}. 
Although some other resonances like $N^*_4(1720), N^*_7(2190)$ contribute to the production but their branching ratios are not known clearly. 

Other channels of $\Lambda$ and $\Sigma$ productions in MB category are $\bar{K}N\rightarrow \Lambda\pi$ and $\bar{K}N\rightarrow \Sigma\pi$, the cross sections of which have been calculated by Ko ~\cite{ko83plb} using K-matrix formalism for three coupled channels $\bar{K}N, \Lambda\pi$ and $\Sigma\pi$. However, we use the experimental parameterized cross section considered in \cite{Brown1} which is in agreement with ~\cite{ko83plb} and is as follows; 
\begin{equation}
\small
\label{eqn_cross_lambdapi}
 \sigma_{{K^-}p\rightarrow \Lambda \pi^0} = \begin{cases}
                                             1.205~p^{-1.428}\text{ mb} & \text{if} ~~p \geq 0.6 \text{ GeV}\\
                                             3.5~p^{0.659}\text{ mb} & \text{if}~~ 0.6 < p \leq 1.0 \text{ GeV}\\
                                             3.5~p^{-3.97}\text{ mb} & \text{if}~~p> 1.0 \text{ GeV}
                                            \end{cases}
\end{equation}
where $p$ in Eq.\ref{eqn_cross_lambdapi} is the anti-Kaon momentum in the laboratory frame. We consider the isospin averaged cross section $\bar{K}N\rightarrow \Lambda\pi$. 

Similarly, the parameterized cross section for $\bar{K}N\rightarrow \Sigma\pi$ is as follows;
\begin{eqnarray}
 \sigma_{{\bar K}N\rightarrow \Sigma \pi}=\sigma_{{K^-}p\rightarrow \Sigma^0 \pi^0}+\sigma_{{K^-}n\rightarrow \Sigma^0 \pi^-}
 \label{eqn_cross_sigmapi}
\end{eqnarray}
where, $\sigma_{{K^-}p\rightarrow \Sigma^0 \pi^0}\approx\sigma_{{K^-}n\rightarrow \Sigma^0 \pi^-}$ and 
\begin{equation}
\footnotesize
 \sigma_{{K^-}p\rightarrow \Sigma^0 \pi^0} = \begin{cases}
                                              0.624~p^{-1.83}\text{ mb} & \text{if} ~~p \leq 0.345 \text{ GeV} \\
                                              \\
                                              \dfrac{0.0138}{(p-0.385)^2+0.0017} \text{ mb} & \text{if}~~ 0.345< p \leq 0.425 \text{ GeV}\\
                                              \\
                                              0.7~p^{-2.09}\text{ mb} & \text{if}~~p> 0.425 \text{ GeV}
                                             \end{cases}
\end{equation}
The contributions from BB category producing single strange hadrons are $pp\rightarrow K\bar{K}$, $pp\rightarrow \Lambda\bar{\Lambda}$, $pp\rightarrow \Sigma\bar{\Sigma}$. The cross-sections for  $p \bar{p}\rightarrow \bar{Y}Y(\bar{M}M)$,(Y is the hyperon, M is the meson, here kaon) is given below ~\cite{kaidalov94,titov08,boreskov83}.
\begin{widetext}
\begin{equation}
 \label{eqn_cross_ppyy}
  \sigma_{P \bar{P}\rightarrow \bar{Y}Y(\bar{K}K)}=
 \frac{C_AC_{Y_{i}(K)}g_{0}^4}{16\pi}\times\frac{s}{s-4m_{P}^{2}}\times\Gamma\left(1-\alpha(0)\right)^2\times
 \left(\frac{s}{s_{0}^{\bar{P}P\rightarrow \bar{Y} Y(\bar{K}K)}}\right)^{2(\alpha_{0}-1)}\times \frac{e^{\varLambda_1 t_{\text{min}}}}{\varLambda_1}
\end{equation}
\end{widetext}
The values of various parameters in the above expression are tabulated in Table~\ref{table_paramet_pp}.
\begin{table}
 \caption{Parameters for $pp\rightarrow YY(MM)$ Reactions}
 \begin{tabular}{ |c|c|c|c|c|c| }
  \hline
   Reactions & $ C_A $ & $ C_{Y_i(M_i)} $ & $\Lambda$ & $s_0$ & Regge   \\
   &  &  & $(GeV^{-2})$ & $(GeV^2)$ & trajectory \\
   &  &  &  & & $\alpha(t)$=  \\
  \hline
   $ pp\rightarrow K^+K^-$ & 0.08 & 4 & 4 & 1.93 & -0.86+0.5t \\   
  \hline
   $ pp\rightarrow \bar{\Lambda}\Lambda$ & 0.10 & 9/4 & 9 & 2.43 &  0.32+0.85t  \\
  \hline
   $ pp\rightarrow \bar{\Sigma^-}\Sigma^+$ & 0.10 & 1 & 9 & 2.43 & 0.32+0.85t  \\
  \hline    
 \end{tabular}
 \label{table_paramet_pp} 
\end{table}
The production cross sections of charged single strange hyperons are 4 times larger than the neutral hyperons from $\bar{p} p$ reactions. It has also been found that 
$\sigma_{p \bar{p}\rightarrow \bar{\Sigma^{-}}\Sigma^{+}}=4\gamma^{4}\sigma_{p \bar{p}\rightarrow \bar{\Lambda}\Lambda}$ and $\sigma_{p \bar{p}\rightarrow \bar{\Lambda}\Lambda}=\frac{9}{4} \sigma_{p \bar{p}\rightarrow \bar{\Sigma^{-}}\Sigma^{+}}$
with $\gamma^2$=1/3. 
The slopes of the differential cross section $\Lambda_1$ for $\Lambda$ and $\Sigma$ are taken to be 9 $GeV^{-2}$ and for $K$ mesons, to be 4 $GeV^{-2}$ by fitting the data on strange hadron production from $\bar{p}-p$ collisions\cite{flaminio84,barnes87,hasan92,tanimori90,sugimoto88}. The value of $g_0$ has been determined from the decay $\rho \rightarrow \pi\pi$ with $\frac{g_{0}^{2}}{4\pi}=2.7$. 

We have considered the inverse reactions  
$K \bar{K} \rightarrow \pi \pi$,
$K \bar{K}\rightarrow \pi \rho $,
$K \bar{K} \rightarrow \rho \rho$,
$K^- \bar{K^+} \rightarrow \bar{p} p$,
$\Lambda K \rightarrow \pi N$, 
$\Lambda K \rightarrow \rho N$, 
$\Sigma K \rightarrow \pi N$, 
$\Lambda \pi \rightarrow \bar{K} N$,
$\Lambda \bar{\Lambda} \rightarrow \bar{p} p$,  
$\Sigma^- \bar{\Sigma^+} \rightarrow \bar{p} p$, 
$\Sigma \pi \rightarrow \bar{K} N$ and the cross sections are calcuated using principle of detailed balance as follows; 
\begin{equation}
 \sigma_{f\rightarrow i}=\frac{{P_i}^2}{{P_f}^2}\frac{g_i}{g_f}\sigma_{i\rightarrow f}
\end{equation}
where $P_i, P_f$ are the cent re of mass momenta and $g_i, g_f$ are the total degeneracies of the initial and final channels. 
Production of single strange hadrons by other channels where multi-strange $\Xi$ and $\Omega$ are involved are described below.  
\subsection{\bf Production cross sections for multi-strange hadrons: cascade($\Xi$) and omega ($\Omega$)}
The multi-strange hadrons are the baryons having strangeness more than one; $\text{S}=\pm 2,\pm 3$. Baryons like Cascade (S=-2) and Omega (S=-3) fall into this category. Due to large strangeness content, the production of multi-strange baryons from non strange hadrons is expensive and less probable. Strangeness exchange reactions become the dominant channels. 
\begin{figure}[H]
\centering
\includegraphics[width=0.45\textwidth,height=50mm]{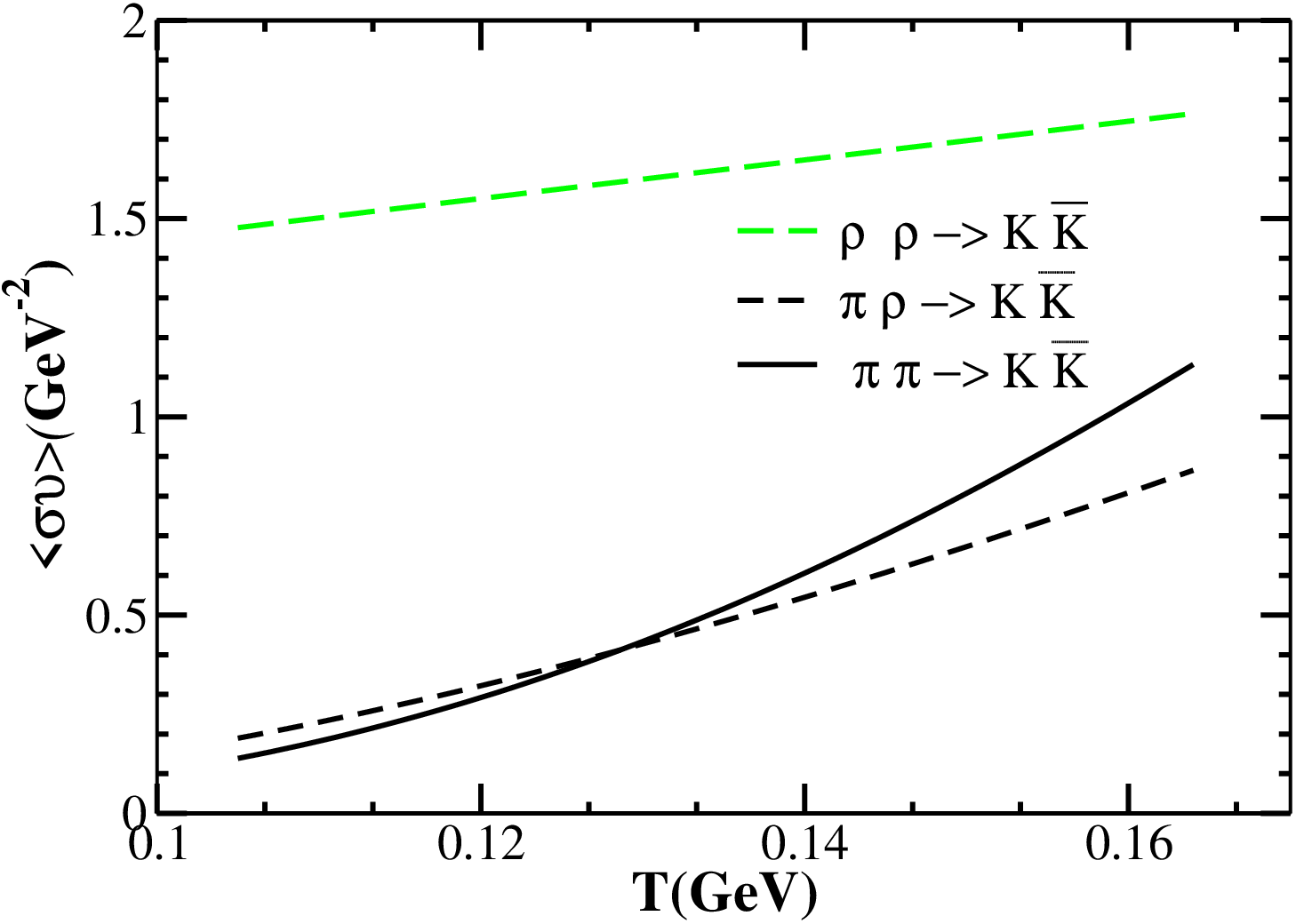}
\caption{Rate (R=$\langle \sigma v\rangle$) of kaon productions from $\pi \pi \rightarrow K\bar{K}$, $\pi \rho \rightarrow K\bar{K}$, $\rho \rho \rightarrow K\bar{K}$ with temperature.}
\label{fig_strange_rate1}
\end{figure}
\begin{figure}[H]
\centering
\includegraphics[width=0.45\textwidth,height=50mm]{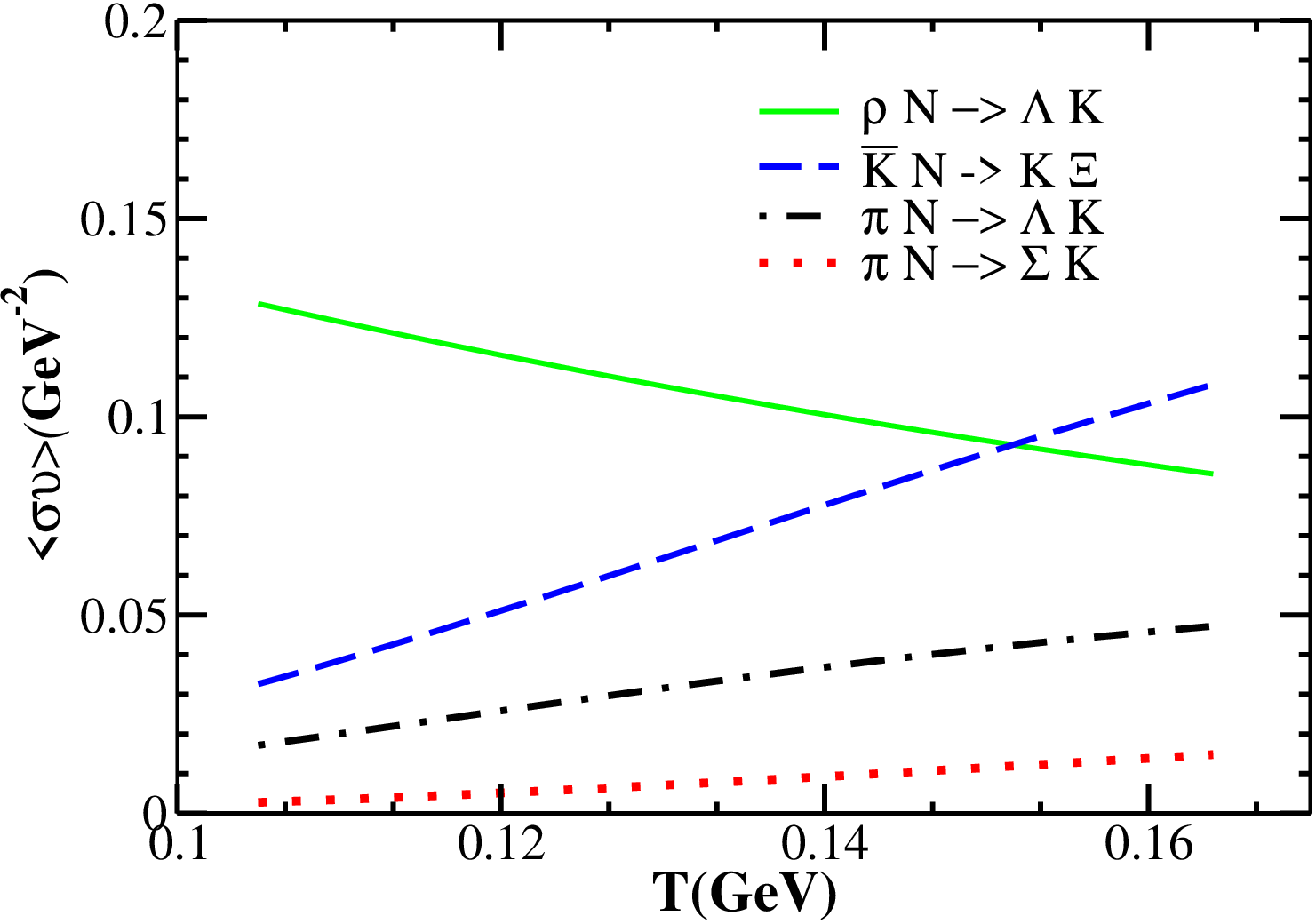}
\caption{Rate (R=$\langle \sigma v\rangle$)from channels $\pi N \rightarrow \Lambda K$, $\rho N \rightarrow \Lambda K$, $\pi N \rightarrow \Sigma K$, $\bar{K} N \rightarrow K \Xi$.}
\label{fig_strange_rate2}
\end{figure}
The types of reactions producing $\Xi(S=-2)$ are $\bar{K}Y\rightarrow \pi\Xi$, $YY\rightarrow B \Xi$, $\bar{K}B\rightarrow K \Xi$ and  $B \bar{B} \rightarrow \Xi \bar{\Xi}$. Here $Y$ represents $\Lambda$ or $\Sigma$. More specifically the reactions are $\Lambda\Lambda\rightarrow N\Xi$, $ \Lambda\Sigma\rightarrow N\Xi$,  $ \Sigma\Sigma\rightarrow N\Xi$, $ \bar{K} \Lambda \rightarrow \pi \Xi$, $ \bar{K} \Sigma \rightarrow \pi \Xi$, $ \bar{K} N \rightarrow K \Xi$, $ \bar{p} p \rightarrow \Xi \bar{\Xi}$. 
\begin{figure*}
\centering
\subfloat{
   \includegraphics[width=0.44\textwidth,height=5.0cm]{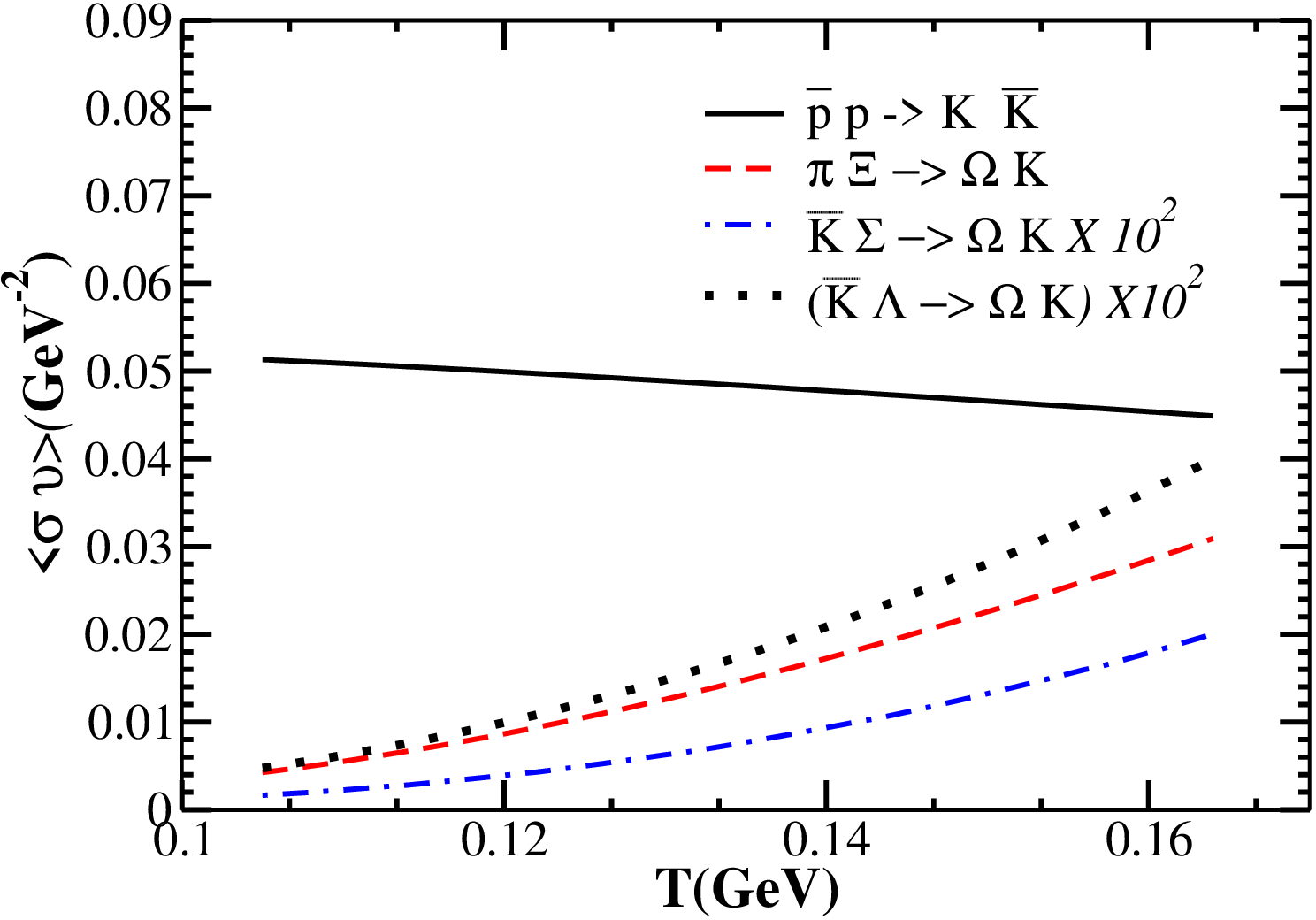}
 }
\subfloat{
   \includegraphics[width=0.44\textwidth,height=5.0cm]{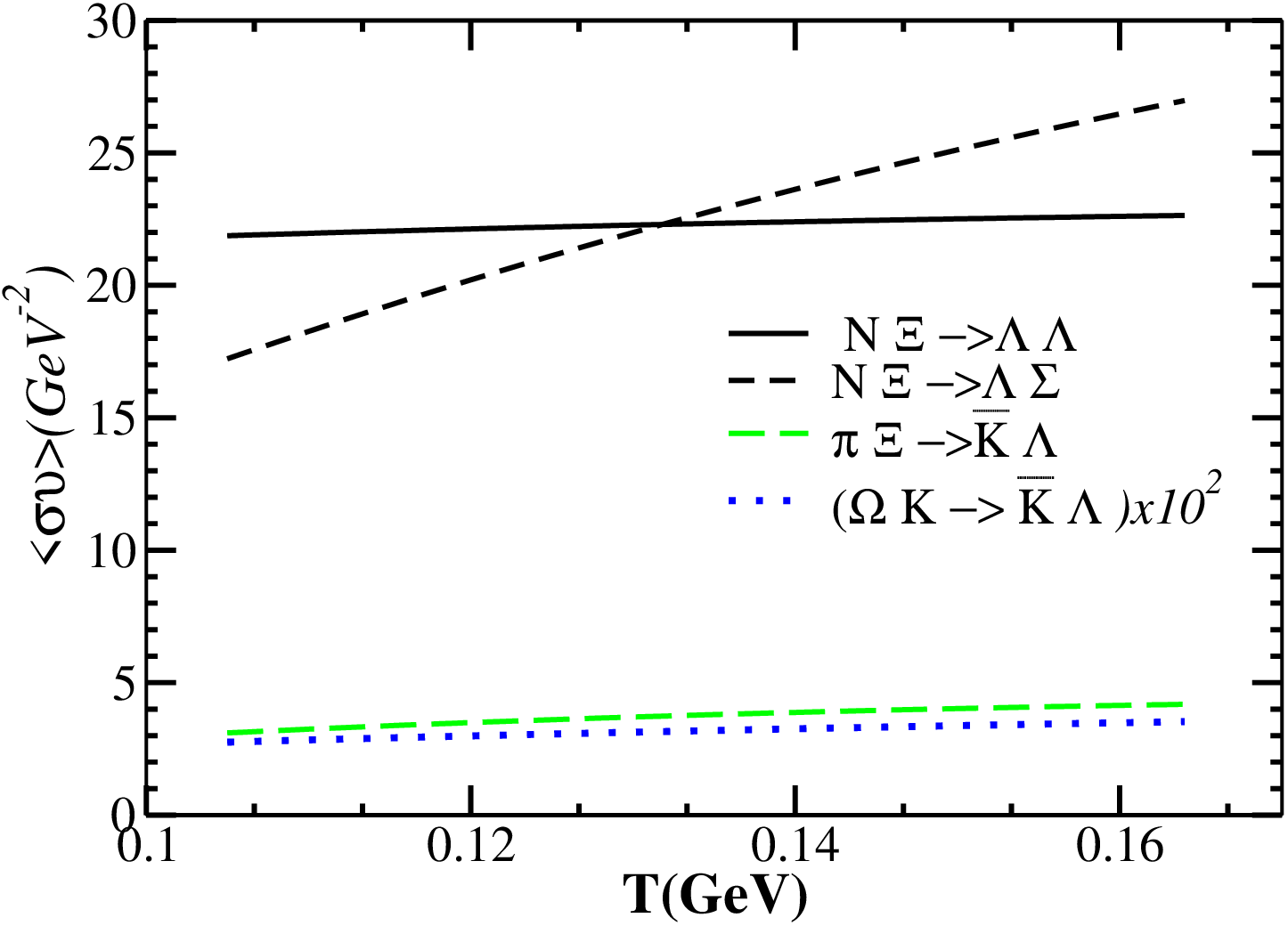}
 }
\caption{Rate ($\langle \sigma v\rangle$) from (left panel:)$\bar{p} p \rightarrow K \bar{K}$, $\pi\Xi \rightarrow \Omega K$, $\bar{K}\Sigma \rightarrow \Omega K$, $\bar{K}\Lambda \rightarrow \Omega K$,(right panel:) $N \Xi \rightarrow \Lambda \Lambda$, $N \Xi \rightarrow \Lambda \Sigma$, $\pi \Xi \rightarrow \bar{K} \Lambda$, $\Omega K \rightarrow K \Lambda$.}
\label{fig_strange_rate3}
\end{figure*}
\begin{figure*}
\centering
\subfloat{
   \includegraphics[width=0.44\textwidth,height=5.0cm]{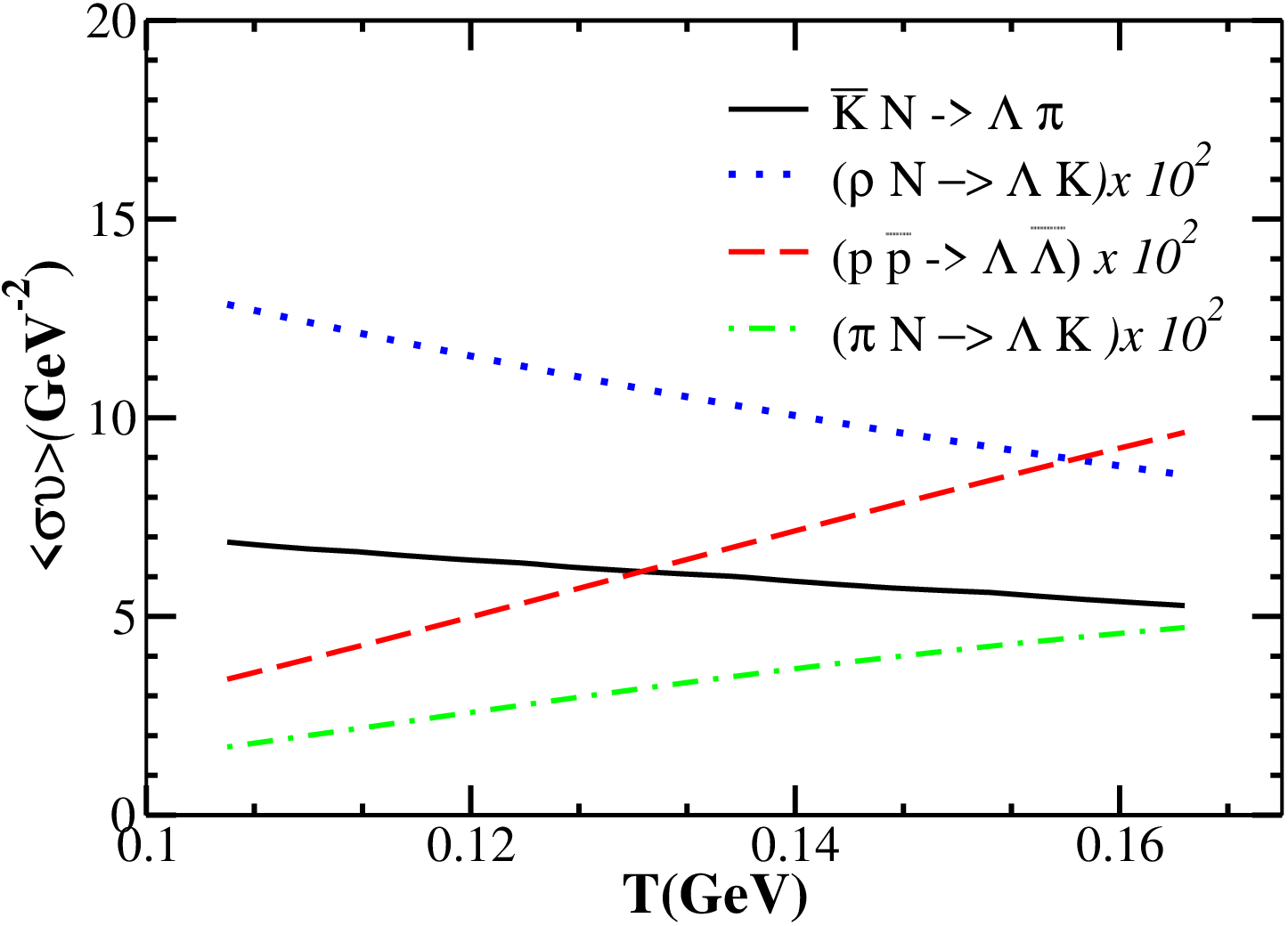}
 }
\subfloat{
   \includegraphics[width=0.44\textwidth,height=5.0cm]{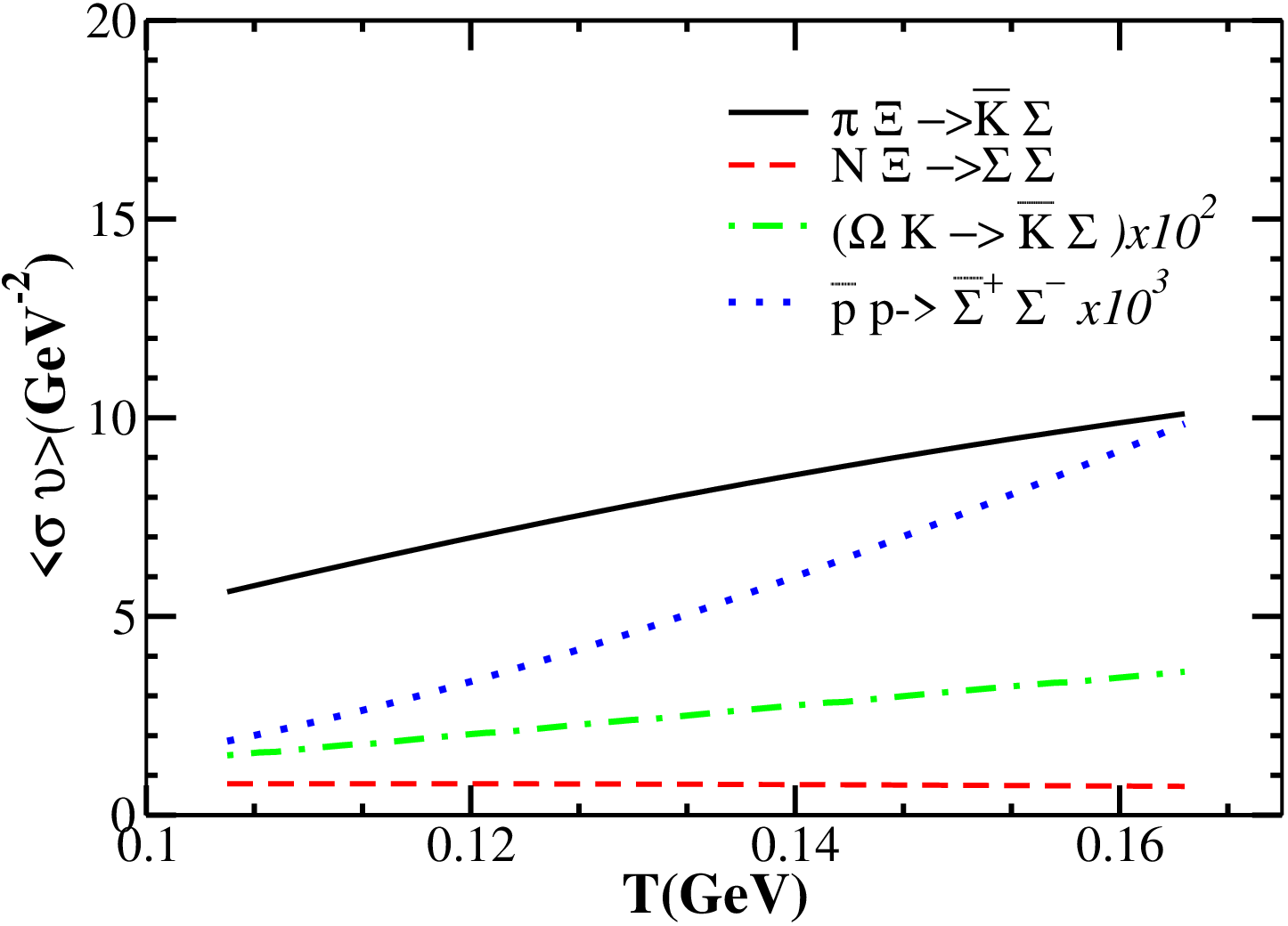}
 }
\caption{Rate (R=$\langle \sigma v\rangle$) from (left panel:) $\bar{K} N \rightarrow \Lambda \pi$ and $p \bar{p}\rightarrow \Lambda \bar{\Lambda}$, (right panel:) $\pi \Xi\rightarrow \bar{K} \Sigma$, $ N \Xi\rightarrow \Sigma \Sigma$,$ \Omega K\rightarrow \bar{K} \Sigma$, $p \bar{p}\rightarrow \Sigma \bar{\Sigma}$.}
\label{fig_strange_rate4}
\end{figure*}
Out of these above mentioned channels, the strangeness exchange reactions are $\Lambda\Lambda\rightarrow N\Xi$, $  \Lambda\Sigma\rightarrow N\Xi$,  $ \Sigma\Sigma\rightarrow N\Xi$, $ \bar{K} \Lambda \rightarrow \pi \Xi$, $ \bar{K} \Sigma \rightarrow \pi \Xi$, $ \bar{K} N \rightarrow K \Xi$. The cross sections are calculated from a SU(3) invariant Langragian density as in~ \cite{liprc12,linpa02}.
\begin{align}
\mathcal{L} &=i\, \text{Tr}\left(\bar{B}\not D B\right)+\text{Tr}\left[D_{\mu} P^{+}D^{\mu}P\right]\nonumber \\
&+g' \, \text{Tr}\left[ \left(2\alpha-1\right)\bar{B}\gamma^5\gamma^{\mu} B D_{\mu} \bar{B}\gamma^5\gamma^{\mu}\left(D_{\mu} P\right)B\right]
 \label{lagrangianeq1}
\end{align}
$B$ and $P$ appearing in the Lagrangian are the baryon and pseudo-scalar meson octets and $D_{\mu}=\partial_{\mu}-ig\left[V_{\mu}\right]$ is the covariant derivative, which accounts for the interaction of pseudo scalar mesons and baryons through pseudo vector($V_{\mu}$) couplings. The octets are 
\addtocounter{table}{-1}
\begin{longtable}{c}
 $B$ =
$\begin{bmatrix}
\frac{\Sigma^0}{\sqrt{2}}+\frac{\Lambda}{\sqrt{6}}& \Sigma^+ & p \\
\Sigma^{-} & \frac{-\Sigma^{0}}{\sqrt{2}}+\frac{\Lambda}{\sqrt{6}} & n \\ 
-\Xi^{-} & \Xi^{0} & -\sqrt{\frac{2}{3}}\Lambda\\
\end{bmatrix}$ \\
$P$ = $\frac{1}{\sqrt{2}}
\begin{bmatrix}
\frac{\pi^{0}}{\sqrt{2}}+\frac{\eta_8}{\sqrt{6}}+\frac{\eta_1}{\sqrt{3}} & \pi^{+} & K^{+}\\
\pi^{-} & \frac{-\pi^0}{\sqrt{2}}+\frac{\eta_8}{\sqrt{6}}+\frac{\eta_1}{\sqrt{3}} & K^0\\
K^{-} & \bar{K}^{0} & -\sqrt{\frac{2}{3}}\eta_8+\frac{\eta_1}{\sqrt{3}}\\
\end{bmatrix}$ \\
$V$ = $\frac{1}{\sqrt{2}}
  \begin{bmatrix}\frac{\rho^0}{\sqrt{2}}+\frac{\omega}{\sqrt{2}} & \rho^+ & K^{*^+}\\
\rho^- & \frac{-\rho^0}{\sqrt{2}}+\frac{\omega}{\sqrt{2}} & K^{*^0}\\
K^{*^-} & \bar{K}^{*^0} & \Phi\\
\end{bmatrix}$
\end{longtable}
The universal coupling constants $g$ and $g'$($g'$ responsible for B-P interactions) are derived from $f_{\pi NN}$, $g_{\rho NN}$ ~\cite{holzenkamp89} and we consider the values $g$=13, $g'$=14.4 GeV and parameter $\alpha$=0.64 \cite{adelseck90}. We also take other relevant couplings from \cite{linpa02}. It has also been found that the contribution of $\eta$ in strangeness exchange reactions is much less compared to the baryons~\cite{linpa02}. Hence we don't consider the interactions of type $\bar{K} \Lambda \rightarrow \eta \Xi$ and $\bar{K} \Sigma \rightarrow \eta \Xi$. 

Tensor interactions, like vector interactions of $V-B$, of D and F types have also been considered by \cite{linpa02,liprc12} and we take the SU(3) invariant Lagrangian
\begin{equation}
 \mathcal{L}^t =\frac{g^t}{2m} \text{Tr}[(2\alpha-1)\bar{B}\sigma^{\mu\nu}B\partial_{\mu}V_{\nu}+\bar{B}\sigma^{\mu\nu}(\partial_\mu V_\nu)B] 
\end{equation}
with $g^t$ obtained from $\rho-N$ tensor coupling~\cite{holzenkamp89} and $m$ is for the degenerate baryon mass. 
The details of the cross section for all these strangeness exchange reactions are calculated in~\cite{liprc12}. We use the parametrised cross section ~\cite{chen04}and evaluate the rate of production which is shown in~\cite{ghosharxiv19}. 

Similarly, $\Xi$ production cross section of $B \bar{B} \rightarrow \Xi \bar{\Xi}$ channel \emph{i.e.} $p \bar{p} \rightarrow \Xi^- \bar{\Xi^+}$ and $p\bar{p}\rightarrow \bar{\Xi}^0\Xi^0$ has been evaluated using quark gluon string model (QGSM)~\cite{kaidalov94}. The results are also compared with experimental observation. Using this cross section the rate has been evaluated in~\cite{ghosharxiv19}. 

We had already discussed that the strangeness exchange channels play crucial role in $\Xi$ productions over  $BB\rightarrow \Xi\bar{\Xi}$. The rates of production of $\Lambda \Lambda \rightarrow N \Xi$ or $\bar{K} \Lambda \rightarrow \pi \Xi$ are $10^6$ times more compared to the channels $p p \rightarrow \Xi \bar{\Xi}$~\cite{ghosharxiv19}. 

Production of $\Omega(S=-3)$ in heavy ion collisions is not well understood.  However we have attempted its study of yield with the current understanding. To mention a few possible reactions for $\Omega$ productions, channels like $\Xi Y \rightarrow \Omega N$ and $\bar{K} \Xi \rightarrow \Omega \pi$ seem to be important as they fall into the category of strangeness exchange reactions. But the production cross sections for these reactions are not clear by now. The authors in \cite{koch89} although argue about its cross section to be similar to $\bar{K} N \rightarrow \pi Y$ but the experimental coupling is not available. Other probable channels we consider are $\pi \Xi \rightarrow \Omega K$, 
($\pi^0 \Xi^- \rightarrow \Omega^- K^0$), $\bar{K} Y \rightarrow K \Omega$ ($\bar{K} \Lambda \rightarrow K^0 \Omega^{-} $, $\bar{K} \Sigma^{0}\rightarrow K^0 \Omega^{-}$) and which are discussed in detail in \cite{ghosharxiv19}. 

The other channel we have considered for $\Omega$ productions is $B\bar{B}\rightarrow \Omega \bar{\Omega}$ or $p\bar{p}\rightarrow \Omega \bar{\Omega}$. The details of cross section and rate of production can be found in \cite{kaidalov94} and \cite{ghosharxiv19}.  

\section{\label{sec:rate} Rate of strange hadron production in hadronic medium}
With the input of cross sections from previous section, the thermal rates of strange hadron productions in hadronic medium are evaluated considering the binary interactions in the following way. The rate, $R(T)$ at a temperature $T$ is given by ~\cite{kapusta86,gondolo91},
\begin{align}
 \langle \sigma v\rangle &=\frac{T^4}{4}\mathcal{C}_{ab}(T)\int _{z_0}^{\infty} \, dz\, [z^2-(m_a/T+m_b/T)^2]\nonumber \\
 &\times [z^2-(m_a/T-m_b/T)^2]\sigma K_1(z)
  \label{eqn_reacrate}
\end{align}
where $\mathcal{C}_{ab}(T)$ is given by
$$\mathcal{C}_{ab}(T) = \frac{1}{m_a^2m_b^2K_2(m_a/T)K_2(m_b/T)}$$
and $\sigma$ is the cross section of particular channel of interest and $v$ is the relative Moller velocity of the incoming particles of masses $m_a$ and $m_b$. $K_2$ is the modified bessel function of second kind. $z_0=\text{max}(m_a+m_b,m_c+m_d)/T$. The detailed derivation of rate and chemical rate equation is given in appendix A. The rate of various channels producing single and multi-strange hadrons are discussed in the result section.
\section{\label{sec:rateequation} Yield of strange hadrons using rate equation}
The number densities of $K, \bar{K}, \Lambda, \Sigma, \Xi$ and $\Omega$ are studied using following rate equations considering the cross sections described in section-\ref{sec:crosssection}. The non-strange mesons and baryons are assumed to provide thermal background to the strange hadrons which are slightly away from equilibrium. Eq.\ref{rate_eqn} describes a set of coupled equations for different strange hadrons and each equation contains terms for net productions due to binary interactions and dilution term ($n_i/t$) due to expansion of the system. 
\begin{widetext}
\begin{align}
\frac{dn_{K}}{dt}+\frac{n_K}{t} &=
n_{\pi}n_{\pi}\langle\sigma v\rangle_{\pi\pi\rightarrow K\bar{K}} 
-n_{K}n_{\bar{K}}\langle\sigma v \rangle_{K\bar{K}\rightarrow \pi\pi} 
+n_{\rho} n_{\rho} \langle\sigma v\rangle_{\rho\rho\rightarrow K\bar{K}} 
-n_{K}n_{\bar{K}}\langle\sigma v\rangle_{K\bar{K}\rightarrow \rho\rho} 
\nonumber\\
& 
+n_{\pi} n_{\rho}\langle\sigma v\rangle_{\pi\rho\rightarrow K\bar{K}}
-n_{K}n_{\bar{K}}\langle\sigma v\rangle_{K\bar{K}\rightarrow \pi\rho}
+n_{\pi} n_N\langle\sigma v\rangle_{\pi N\rightarrow \Lambda K}
-n_{\Lambda} n_K\langle\sigma v\rangle_{\Lambda K\rightarrow \pi N}
\nonumber\\
& 
+n_{\rho} n_N\langle\sigma v\rangle_{\rho N\rightarrow \Lambda K}
-n_{\Lambda} n_K\langle\sigma v\rangle_{\Lambda K\rightarrow \rho N}
+n_{\pi} n_N\langle\sigma v\rangle_{\pi N\rightarrow \Sigma K}
-n_{\Sigma} n_K\langle\sigma v\rangle_{\Sigma K\rightarrow \pi N}
\nonumber\\
& 
+n_{\bar K} n_N\langle\sigma v\rangle_{\bar{K}N\rightarrow K\Xi}
 -n_K n_{\Xi}\langle\sigma v\rangle_{K\Xi\rightarrow \bar{K}N}
+n_p n_{\bar{p}} \langle\sigma v\rangle_{ p \bar{p}\rightarrow K \bar{K}}
-n_K n_{\bar{K}}\langle\sigma v\rangle_{K \bar{K} \rightarrow p \bar{p}}
\nonumber\\ 
&
+n_{\bar{K}} n_{\Lambda} \langle\sigma v\rangle_{\bar{K} \Lambda\rightarrow \Omega K}
-n_{\Omega}n_{K}\langle\sigma v\rangle_{\Omega K \rightarrow \bar{K}\Lambda}
+n_{\bar{K}} n_{\Sigma} \langle\sigma v\rangle_{\bar{K} \Sigma\rightarrow \Omega K}
-n_{\Omega}n_{K}\langle\sigma v\rangle_{\Omega K \rightarrow \bar{K}\Sigma}
\nonumber\\
&
+n_{\pi} n_{\Xi}\langle\sigma v\rangle_{\pi \Xi \rightarrow K\Omega}
-n_{\Omega} n_{K}\langle\sigma v\rangle_{ \Omega K \rightarrow \pi \Xi} \nonumber \\
\frac{dn_{\bar{K}}}{dt} +\frac{n_{\bar{K}}}{t} &= 
n_{\pi}n_{\pi}\langle\sigma v\rangle_{\pi\pi\rightarrow K\bar{K}} 
-n_{K}n_{\bar{K}}\langle\sigma v \rangle_{K\bar{K}\rightarrow \pi\pi}
+n_{\rho} n_{\rho} \langle\sigma v\rangle_{\rho\rho\rightarrow K\bar{K}} 
-n_{K}n_{\bar{K}}\langle\sigma v\rangle_{K\bar{K}\rightarrow \rho\rho} 
\nonumber\\
&
+n_{\pi} n_{\rho}\langle\sigma v\rangle_{\pi\rho\rightarrow K\bar{K}}
-n_{K}n_{\bar{K}}\langle\sigma v\rangle_{K\bar{K}\rightarrow \pi\rho}
-n_{\bar{K}} n_N \langle\sigma v\rangle_{\bar{K}N\rightarrow \Lambda \pi}
+n_{\Lambda}n_{\pi}\langle\sigma v\rangle_{\Lambda \pi\rightarrow \bar{K}N}
\nonumber\\
&
-n_{\bar{K}} n_N \langle\sigma v\rangle_{\bar{K}N\rightarrow \Sigma\pi}
+n_{\Sigma}n_{\pi}\langle\sigma v\rangle_{\Sigma \pi\rightarrow \bar{K}N}
-n_{\bar K} n_N\langle\sigma v\rangle_{\bar{K}N\rightarrow K\Xi}
+n_K n_{\Xi}\langle\sigma v\rangle_{K\Xi\rightarrow \bar{K}N}
\nonumber\\
&
-n_{\bar{K}} n_{\Lambda}\langle\sigma v\rangle_{\bar{K}\Lambda\rightarrow \pi\Xi}
+n_{\pi} n_{\Xi}\langle\sigma v\rangle_{\pi\Xi \rightarrow \bar{K}\Lambda}
-n_{\bar{K}} n_{\Sigma}\langle\sigma v\rangle_{\bar{K}\Sigma\rightarrow \pi\Xi}
+n_{\pi} n_{\Xi}\langle\sigma v\rangle_{\pi\Xi \rightarrow \bar{K}\Sigma}
\nonumber\\
&
+n_p n_{\bar{p}} \langle\sigma v\rangle_{ p\bar{p} \rightarrow  K \bar{K}}
-n_{K}n_{\bar{K}}\langle\sigma v\rangle_{ K \bar{K} \rightarrow p \bar{p}}
-n_{\bar{K}} n_{\Lambda} \langle\sigma v\rangle_{\bar{K} \Lambda\rightarrow \Omega K}
+n_{\Omega}n_{K}\langle\sigma v\rangle_{\Omega K \rightarrow \bar{K}\Lambda}
\nonumber\\
&
-n_{\bar{K}} n_{\Sigma} \langle\sigma v\rangle_{\bar{K} \Sigma\rightarrow \Omega K}
+n_{\Omega}n_{K}\langle\sigma v\rangle_{\Omega K \rightarrow \bar{K}\Sigma}
 \nonumber \\
\frac{dn_{\Lambda}}{dt}+\frac{n_{\Lambda}}{t}&=
n_{\pi} n_N\langle\sigma v\rangle_{\pi N\rightarrow \Lambda K}
-n_{\Lambda} n_K\langle\sigma v\rangle_{\Lambda K\rightarrow \pi N}
+n_{\rho} n_N\langle\sigma v\rangle_{\rho N\rightarrow \Lambda K}
-n_{\Lambda} n_K\langle\sigma v\rangle_{\Lambda K\rightarrow \rho N}
\nonumber\\
&
-n_{\Lambda} n_{\Lambda}\langle\sigma v\rangle_{\Lambda\Lambda\rightarrow N\Xi}
+n_N n_{\Xi}\langle\sigma v\rangle_{N \Xi\rightarrow \Lambda\Lambda}
-n_{\Lambda} n_{\Sigma}\langle\sigma v\rangle_{\Lambda\Sigma\rightarrow N\Xi}
+n_N n_{\Xi}\langle\sigma v\rangle_{N \Xi\rightarrow \Lambda\Sigma}
\nonumber\\
&
-n_{\bar K} n_{\Lambda}\langle\sigma v\rangle_{\bar{K}\Lambda\rightarrow \pi\Xi}
+n_{\pi} n_{\Xi}\langle\sigma v\rangle_{\pi\Xi \rightarrow \bar{K}\Lambda}
+n_{\bar K} n_{N}\langle\sigma v\rangle_{\bar{K}N\rightarrow {\Lambda}\pi}
-n_{\Lambda} n_{\pi}\langle\sigma v\rangle_{\Lambda\pi \rightarrow \bar{K}N}
\nonumber\\
&
+n_p n_{\bar{p}}\langle\sigma v\rangle_{p \bar{p}\rightarrow \Lambda\bar{\Lambda}}
-n_{\Lambda} n_{\bar{\Lambda}}\langle\sigma v\rangle_{ \Lambda \bar{\Lambda}\rightarrow p\bar{p}} 
+n_{K} n_{\Omega}\langle\sigma v\rangle_{K{\Omega} \rightarrow {\bar K}{\Lambda}}
-n_{{\bar K}} n_{\Lambda}\langle\sigma v\rangle_{{\bar K} {\Lambda}\rightarrow K \Omega} \nonumber\\
\frac{dn_{\Sigma}}{dt}+ \frac{n_{\Sigma}}{t}&=
n_{\pi} n_N\langle\sigma v\rangle_{\pi N\rightarrow \Sigma K}
-n_{\Sigma} n_K\langle\sigma v\rangle_{\Sigma K\rightarrow \pi N}
-n_{\Lambda} n_{\Sigma}\langle\sigma v\rangle_{\Lambda\Sigma\rightarrow N\Xi}
+n_N n_{\Xi}\langle\sigma v\rangle_{N \Xi\rightarrow \Lambda\Sigma}
\nonumber\\
&
-n_{\Sigma} n_{\Sigma}\langle\sigma v\rangle_{\Sigma\Sigma\rightarrow N\Xi}
+n_N n_{\Xi}\langle\sigma v\rangle_{N \Xi\rightarrow \Sigma\Sigma}
-n_{\bar{K}} n_{\Sigma}\langle\sigma v\rangle_{\bar{K}\Sigma\rightarrow \pi\Xi}
 +n_{\pi} n_{\Xi}\langle\sigma v\rangle_{\pi\Xi \rightarrow \bar{K}\Sigma} 
\nonumber\\
&
+n_{\bar K} n_{N}\langle\sigma v\rangle_{\bar{K}N\rightarrow {\Sigma}\pi}
-n_{\Sigma} n_{\pi}\langle\sigma v\rangle_{\Sigma\pi \rightarrow \bar{K}N}
+n_p n_{\bar{p}}\langle\sigma v\rangle_{p \bar{p}\rightarrow \Sigma\bar{\Sigma}}
-n_{\Sigma} n_{\bar{\Sigma}}\langle\sigma v\rangle_{ \Sigma \bar{\Sigma}\rightarrow p\bar{p}}
\nonumber\\
&
+n_{K} n_{\Omega}\langle\sigma v\rangle_{K{\Omega} \rightarrow {\bar K}{\Sigma}}
-n_{{\bar K}} n_{\Sigma}\langle\sigma v\rangle_{{\bar K} {\Sigma}\rightarrow K \Omega} \nonumber \\
\frac{dn_{\Xi}}{dt}+\frac{n_{\Xi}}{t}&=
n_{\Lambda} n_{\Lambda}\langle\sigma v\rangle_{\Lambda\Lambda\rightarrow N\Xi}
-n_N n_{\Xi}\langle\sigma v\rangle_{N \Xi\rightarrow \Lambda\Lambda}
+n_{\Lambda} n_{\Sigma}\langle\sigma v\rangle_{\Lambda\Sigma\rightarrow N\Xi}
-n_N n_{\Xi}\langle\sigma v\rangle_{N \Xi\rightarrow \Lambda\Sigma}
\nonumber\\
&
+n_{\Sigma} n_{\Sigma}\langle\sigma v\rangle_{\Sigma\Sigma\rightarrow N\Xi}
-n_N n_{\Xi}\langle\sigma v\rangle_{N \Xi\rightarrow \Sigma\Sigma}
+n_{\bar K} n_N\langle\sigma v\rangle_{\bar{K}N\rightarrow K\Xi}
-n_K n_{\Xi}\langle\sigma v\rangle_{K\Xi\rightarrow \bar{K}N}
\nonumber\\
& 
+n_{\bar{K}} n_{\Lambda}\langle\sigma v\rangle_{\bar{K}\Lambda\rightarrow \pi\Xi}
-n_{\pi} n_{\Xi}\langle\sigma v\rangle_{\pi\Xi \rightarrow \bar{K}\Lambda}
+n_{\bar{K}} n_{\Sigma}\langle\sigma v\rangle_{\bar{K}\Sigma\rightarrow \pi\Xi}
-n_{\pi} n_{\Xi}\langle\sigma v\rangle_{\pi\Xi \rightarrow \bar{K}\Sigma}\nonumber\\
&+n_p n_{\bar{p}}\langle\sigma v\rangle_{p \bar{p}\rightarrow \Xi\bar{\Xi}}
-n_{\Xi} n_{\bar{\Xi}}\langle\sigma v\rangle_{ \Xi \bar{\Xi}\rightarrow p\bar{p}} 
+n_{\Omega} n_{K}\langle\sigma v\rangle_{ \Omega K\rightarrow {\pi}{\Xi}}-
n_{\pi} n_{\Xi}\langle\sigma v\rangle_{\pi {\Xi}\rightarrow \Omega K} \nonumber\\
\frac{dn_{\Omega}}{dt}+ \frac{n_{\Omega}}{t} &=
n_p n_{\bar{p}}\langle\sigma v\rangle_{p \bar{p}\rightarrow \Omega\bar{\Omega}}
-n_{\Omega} n_{\bar{\Omega}}\langle\sigma v\rangle_{ \Omega \bar{\Omega}\rightarrow p\bar{p}}
+n_{\pi} n_{\Xi}\langle\sigma v\rangle_{\pi {\Xi}\rightarrow \Omega K}
-n_{\Omega} n_{K}\langle\sigma v\rangle_{ \Omega K\rightarrow {\pi}{\Xi}} 
\nonumber\\
&
+n_{{\bar K}} n_{\Lambda}\langle\sigma v\rangle_{{\bar K} {\Lambda}\rightarrow K \Omega}
-n_{K} n_{\Omega}\langle\sigma v\rangle_{ K\Omega\rightarrow{\bar K}{\Lambda}} 
+n_{\bar K} n_{\Sigma}\langle\sigma v\rangle_{{\bar K} {\Sigma}\rightarrow K \Omega}
-n_{K} n_{\Omega}\langle\sigma v\rangle_{K\Omega\rightarrow{\bar K}{\Sigma}}
 \label{rate_eqn}
\end{align}
\end{widetext}
\begin{figure*}
\centering
\subfloat{
   \includegraphics[width=0.44\textwidth,height=5.0cm]{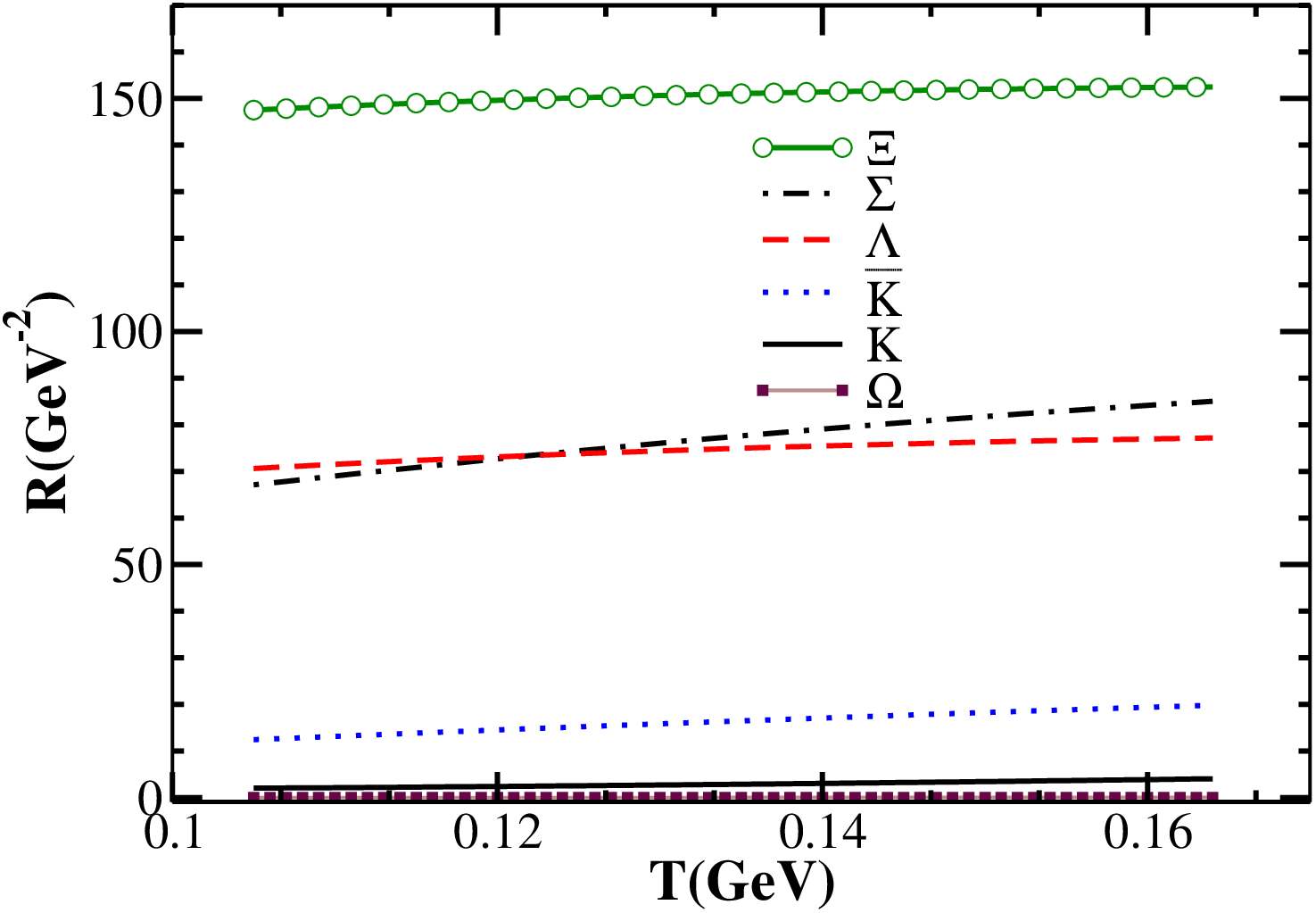}
 }
\subfloat{
   \includegraphics[width=0.44\textwidth,height=5.0cm]{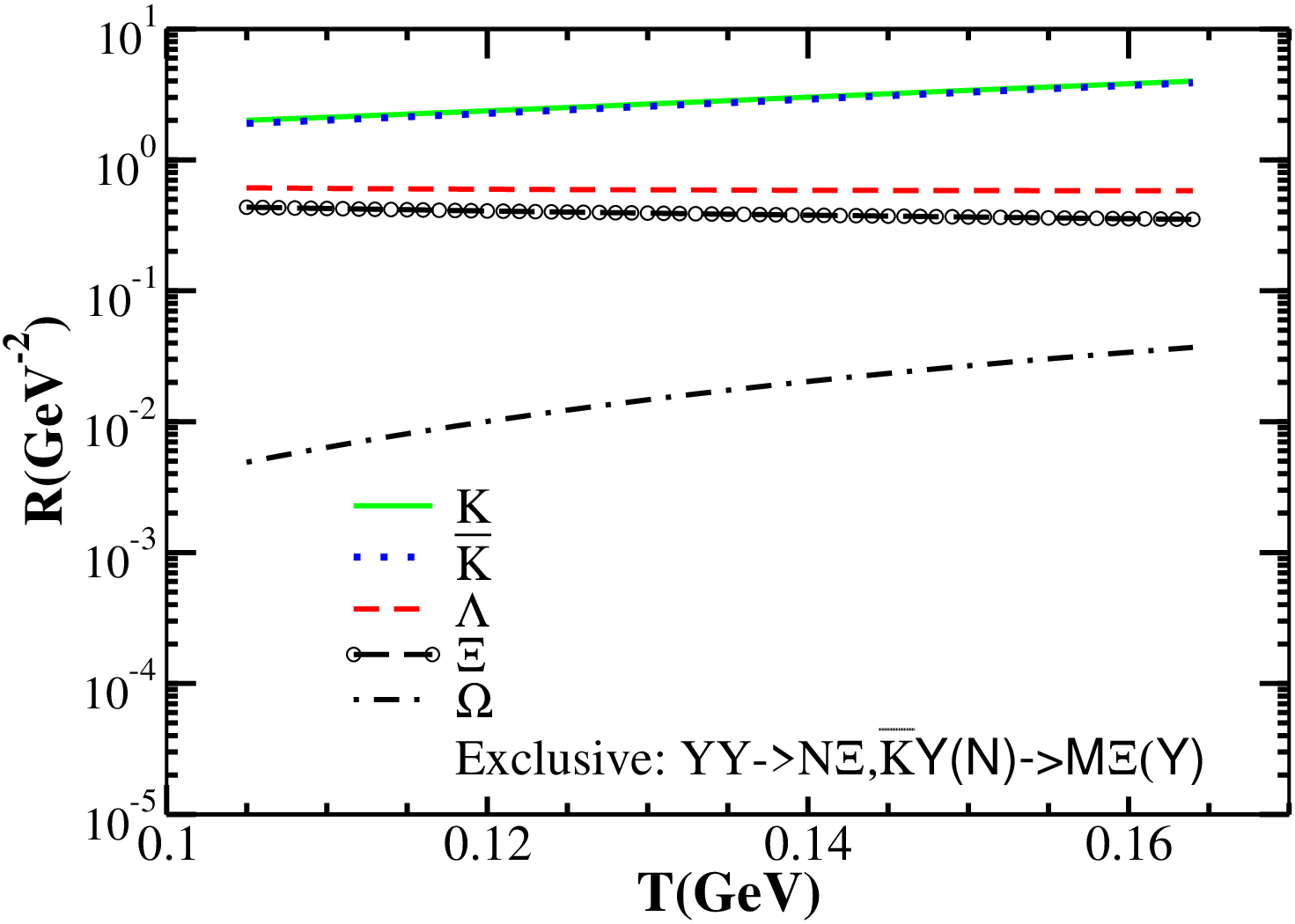}
 }
\caption{Rate ($\langle \sigma v\rangle$) from (left panel:)Total rates (R=$\langle \sigma v\rangle$)of $K, \Lambda, \Sigma, \Xi$ and $\Omega$ production. Top panel is obtained with all reaction channels mentioned in the section\ref{sec:rate},(right panel:) Total rates (R=$\langle \sigma v\rangle$)of $K, \Lambda, \Sigma, \Xi$ and $\Omega$ production excluding the cascade production channels $YY \rightarrow N\Xi$, $\bar{K}Y(N)\rightarrow \pi(K)\Xi$ and the inverse channels producing $\Lambda$ and $\Sigma$. The rates in the right panel are in log scale.}
\label{fig_strange_rate_comp}
\end{figure*}
We do not consider initial QGP phase in this study. The information of the strange production from QGP phase should , in principle, constrain the initial number densities($n_i(T_i)$) of the rate equations in hadronic phase, where $T_i$ is the initial temperature. To take care of this we treat $n_i(T_i)$ as parameters here. These rate equations are numerically coded as Strange Hadron Transport in Heavy Ion Collisions(SH-THIC) to get the yield along with temperature evolution equation considering Bjorken expansion of the system. Although present study is for LHC energy, $\sqrt{s_{NN}}$=2.76 TeV Pb-Pb collisions, where the baryonic chemical potential($\mu_b$) is very small, still we have considered the evolution of $\mu_b$ for the sake of completeness.

The evolution of the number density depends on the evolution 
of the temperature and chemical potential $\mu$ ($=\mu_s+\mu_b$ ). We consider net $\mu_s$ to be zero and $\mu=\mu_b$ is the total chemical potential. 
When we collide two nuclei in heavy ion collision the net strange content is zero. This suggests to assume zero strangeness chemical potential for the produced system from strangeness conservation. However, strangeness chemical potential is also 
related to baryon chemical potential or net baryons in the system. Conservation of baryon number may lead to small strangeness potential. Since we are analysing the matter produced at LHC, the baryonic chemical potential here is very small, hence it is good to assume strange chemical potential to be zero. The evolution of baryonic chemical potential is obtained from the baryon number conservation equation with Bjorken expansion along z- direction as follows,
\begin{equation}
\partial_{\mu}n_b^{\mu}=0
\end{equation}
where, $n_b^{\mu}=n_b u^{\mu}=n_b(\gamma,0,0, \gamma {v_z})$ with $n_b$ is the net baryon number density at ($T,\mu$). The above equation leads to $n_b\tau$=\text{const.}=$k_1$ and $n_b=\sum_{B=N,\Lambda,\Sigma,\Xi,\Omega}(n_B-n_{\bar B})$ and $\tau$ is the proper time defined by $\tau=\sqrt{t^2-z^2}$. The evolution of $\mu_b$ is obtained from the above equation. We have not considered $\Delta$ and other massive baryons as contribution is less due to mass. Again, following Bjorken expansion~\cite{bjorken} and energy conservation law 
$\partial_{\mu}T^{\mu\nu}=0$, we get, $\frac{\partial}{\partial\tau}\left[{T^{\frac{4}{(1+c_s^2)}}\tau}\right]=0$ or $T^a\tau=\text{const.}=k_2$ and $a=\frac{4}{(1+c_s^2)}$ considering energy density $\epsilon$ that goes as $\sim T^4$. As usual the $T^{\mu\nu}$ represents the energy momentum tensor of the expanding fluid. $c_s^2$ is the square of the velocity of sound. Here $k_1=n_b^i\tau_i$, $k_2= T_i^a \tau_i$, where $n_b^i, \tau_i, T_i$ are the initial baryon number densities, time and temperature and are parametres. $T_i$ is taken as the $T_c$ from the lattice calculation. 
\begin{table*}
 \centering 
 \caption{Initial conditions (Freeze out temperatures, $T_F$) for various multiplicities of $K_s^0$, $\Lambda$, $\Xi$ and $\Omega$. At multiplicities 1601 and 1294, 13.4 kaon and lambda data are only available. At multiplicity 1447.5, cascade and omega data are available only.} 
 \begin{tabular}{ |c|c|c|c|c|c|c| } 
  \hline
  $dn_{ch}/d\eta$ & $N_{part}$ & $c_s^{2}$ & Scenario-I & Scenario-II & Scenario-III & Scenario-IV \\
   & & & $T_{f_{1}}$ & $ T_{f_{2}}$ & $T_{f_{3}}$ & $T_{f_{4}}$  \\
    & & & (in GeV) & (in GeV) & (in GeV) & (in GeV)  \\
    \hline
   1601 & 383 & 1/5 & 0.152($K^0_s,\Lambda$) & 0.144 & 0.144 & 0.154  \\
  \hline
   1447.5 & 356.1 & 1/5 & 0.148 ($\Xi,\Omega$) & 0.144 & 0.144 & 0.154  \\ 
   \hline
   1294 & 330 & 1/5 & 0.148 ($K^0_s,\Lambda$) & 0.144 & 0.144 & 0.154  \\
  \hline
   966 & 260.1 & 1/5  & 0.145 & 0.144 & 0.144 & 0.154   \\
  \hline
   537.5 & 157.2 & 1/5 & 0.141 & 0.144 & 0.144 & 0.154  \\
  \hline
   205 & 68.6 & 1/5 & 0.130 & 0.144 & 0.144 & 0.154  \\
  \hline
   55 & 22.5 & 1/5 & 0.114 & 0.144 &  0.144  & 0.154  \\
  \hline
   13.4 & 4.3 & 1/5 & 0.100($K^0_s,\Lambda$) & 0.144 & 0.144 & 0.154  \\
  \hline
 \end{tabular}
 \label{table_scenario_1234} 
\end{table*}
\begin{table*}
 \caption{Initial conditions with freeze out temperature for scenario-V, that explains the data. The * symbol says about the unavailability of data at those multiplicities.}
 \centering
 \begin{tabular}{ |c|c|c|c|c|c|c| }
   \hline
  $dn_{ch}/d\eta$ & $N_{part}$ & ${C_s}^{2}$ & Scenario-V & Scenario-V & Scenario-V & Scenario-V \\
   & & & $T_{f_{5}}(K_s^0)$ & $ T_{f_{5}}(\Lambda)$ & $T_{f_{5}}(\Xi)$ & $T_{f_{5}}(\Omega)$  \\
    & & & (in GeV) & (in GeV) & (in GeV) & (in GeV)  \\
   \hline
   1601 & 383  & 1/5 & 0.154 & 0.156 & * &*\\
   \hline
   1294 & 330  & 1/5 & 0.153 & 0.156 & *&*\\
   \hline
   1447.5 & 356.1 & 1/5 & *  & *  & 0.155  & 0.156    \\ 
  \hline
   966 & 260.1 & 1/5  & 0.153  & 0.155 & 0.156 & 0.156    \\
  \hline
   537.5 & 157.2 & 1/5 & 0.152 & 0.154 & 0.156  & 0.156  \\
  \hline
   205 & 68.6 & 1/5 & 0.150 & 0.151 & 0.154 & 0.154 \\
  \hline
   55 & 22.5 & 1/5 & 0.146 & 0.146 & 0.146 & 0.146    \\
  \hline
  13.4 & 4.3 & 1/5 & 0.141 & 0.141 &* &*  \\
  \hline
 \end{tabular}
 \label{table_scenario5} 
 \end{table*}
After solving the rate equations with the evolution of temperature and chemical potential the yields have been calculated and discussed in the next section.
\section{\label{sec:results} Results}
Taking the cross sections from earlier section as input, the rate of production (R=$\langle \sigma v\rangle _{ab\rightarrow cd}$) for strange hadrons $K, \bar{K}, \Lambda, \Sigma, \Xi$ and $\Omega$ have been calculated from Eq.\ref{eqn_reacrate}. The rates have been displayed in Figs.\ref{fig_strange_rate1}-\ref{fig_strange_rate_comp} for the temperature ranges of our interest. Here we describe the rate of single strange hadrons ($K, \Lambda, \Sigma$) more explicitly as the multi strange hadron rates and yields are described in~\cite{ghosharxiv19} in detail. However the total production rates of $\Xi$ and $\Omega$ are discussed later.

The rate of Kaon($K,\bar{K}$) productions from meson-meson(MM) interactions are shown in Fig.~\ref{fig_strange_rate1} for a temperature range  105-170 MeV. The rate increases with temperature as expected. We have considered only binary interactions for strange hadron productions. Among these binary channels $\rho\rho \rightarrow K\bar{K}$ is the dominant one. $\pi\pi \rightarrow K\bar{K}$ and $\pi\rho \rightarrow K\bar{K}$ have similar contributions over the entire range of temperature as shown in the figure. Fig.~\ref{fig_strange_rate2} shows the rate of Kaon productions along with hyperon($\Lambda,\Sigma~ \& ~\Xi$) productions from meson-baryon(M-B) interactions. Contrary to the increase of rate with temperature, $\rho N$ channel shows a gradual decrease which is due the behaviour of cross section with centre of mass energies of the colliding $\rho$ and $N$ in the thermal system within the considered temperature range. $\rho N$ channel dominates over other channels in this($MB$) category  when the system is at lower temperature. 

Similarly, other process producing Kaons ($ K,\bar{K}$) is the interaction of $p-\bar{p}$ which is shown in the left panel of Fig.~\ref{fig_strange_rate3}. Kaons are also produced from strangeness exchange reactions along with $\Xi$ and $\Omega$. Basically $\pi\Xi \rightarrow \Omega K$, $\bar{K}\Sigma \rightarrow \Omega K$, $\bar{K}\Lambda \rightarrow \Omega K$ are the channels, whose contributions are less to the kaon production but important for $\Omega$ productions, which are shown in Fig.\ref{fig_strange_rate3}. In the right panel of the Fig.\ref{fig_strange_rate3}, the $\Lambda$ production rates are shown from strangeness exchange reactions.  $K$ or $\Sigma$ are the associated particles in the out going channel. These channels play dominant role for the yield of light hyperons $\Lambda$ and $\Sigma$. The cross sections of $N \Xi \rightarrow \Lambda \Lambda$ and $N \Xi \rightarrow \Lambda \Sigma$ are most crucial for the $\Lambda$ productions. However we have excluded these processes because of unreasonable production cross sections of the inverse processes (producing $\Xi$) and there is no experimental verification. Another process which involves K and $\Lambda$ productions is $\Omega K \rightarrow \bar{K}\Lambda$. Contribution from this channel is less due to the massive $\Omega$ in the initial channel as shown in the right panel of Fig.\ref{fig_strange_rate3}.

Rates from $MB$ ($\bar{K} N, \rho N,\pi N$) and $BB(pp)$ interactions producing $K, \bar{K}$, and $\Lambda$ are shown in the left panel of Fig.\ref{fig_strange_rate4}. These processes have negligible contributions compared to $\bar{K}N\rightarrow \Lambda\pi$. The rates of $\Sigma$ production can also be understood from Figs.\ref{fig_strange_rate2}, \ref{fig_strange_rate3} and Fig.\ref{fig_strange_rate4}. 

As far as the production rate of $\Xi$ is concerned the possible processes with initial channel $YY, KY, \bar{K}N$ and $N\bar{N}$ are already discussed. However, we don't consider $YY\rightarrow N\Xi$ and $\bar{K} N\rightarrow \pi\Xi$ for the net yield and the reason is mentioned in the paragraph below. In fact, the contribution from $YY$ channel is dominant and decide the cascade production. The variation of rate with temperature is slow. For details of rate of $\Xi,\Omega$ productions, one can see \cite{ghosharxiv19}. The total rates of $K, \bar{K}, \Lambda, \Xi$ and $\Omega$ are shown and compared in Fig.\ref{fig_strange_rate_comp}.  

From the above figures it is observed that the rate of $\Xi$ production is more than that of  kaon and $\Lambda$; the rate of $\Lambda$ production is more than $K$. It is not expected. It was found that the higher cascade production rate is because of reaction channels $YY \rightarrow N\Xi$, $\bar{K}Y(N)\rightarrow \pi(K)\Xi$ which are calculated using a  Lagrangian in Li etal.\cite{liprc12,linpa02}. Theoretical crosssections for these channels are not constrained experimentally. The inverse process that produces $\Lambda$ along with $\bar{K} N\rightarrow \pi \Lambda$, in fact,  increases the rate of $\Lambda$ production than the kaon. Hence we don't consider these channels of $\Xi$ and $\Lambda$ productions. Without these channels we get the rate of production of $K$ to be more than $\Lambda$ and $\Lambda$ to be more than $\Xi$ as expected. Fig.\ref{fig_strange_rate_comp} displays the total rate of production of all strange hadrons with and without the above mentioned channels.  

Now onwards we have excluded these two types of processes related to $\Xi$ production and the inverse channels. 

Then we solve the rate equations simultaneously to get the number densities of $K, \bar{K}, \Lambda, \Xi, \Omega$ with various initial conditions. Various scenarios are mentioned below and the parameters of initial conditions are tabulated in Table-\ref{table_scenario_1234} and Table-\ref{table_scenario5}. The number densities are then normalised with thermal pion number density to obtain the {\it yield ratio}.

We consider the following scenarios as mentioned below. 
\begin{itemize}
 \item 
 Scenario-I: initial number densities of strange hadrons are assumed to be 15\% away from the equilibrium value i.e., $n_i=0.85 n_{eq}(T_i)$. Freest temperatures decrease with centrality.
 \item
 Scenario-II: 
 initial density is 40\% away from the equilibrium value with constant freeze out temperature 144 MeV for all centralities.
 \item
 Scenario-III:
 initial number density is 15\% away from equilibrium with constant freeze out temperature 144 MeV.
 \item
 Scenario-IV:
 initial number density is 40\% away from equilibrium with constant freeze out temperature is 154 MeV (motivated from Statistical Hadronisation Model at 2.76 TeV LHC energy\cite{alice_adam2016}).
 \item
 Scenario-V:
 initial number density is 15\% away from the equilibrium value and considered the temperature that best explain all the data simultaneously. 
 \item
 Scenario-VI:
 initial number density is 15\% away from equilibrium and $T_F$=154 MeV constant for all centralities.
 \end{itemize}

\begin{figure}
\centering
\subfloat{\includegraphics[width=0.42\textwidth,height=5cm]{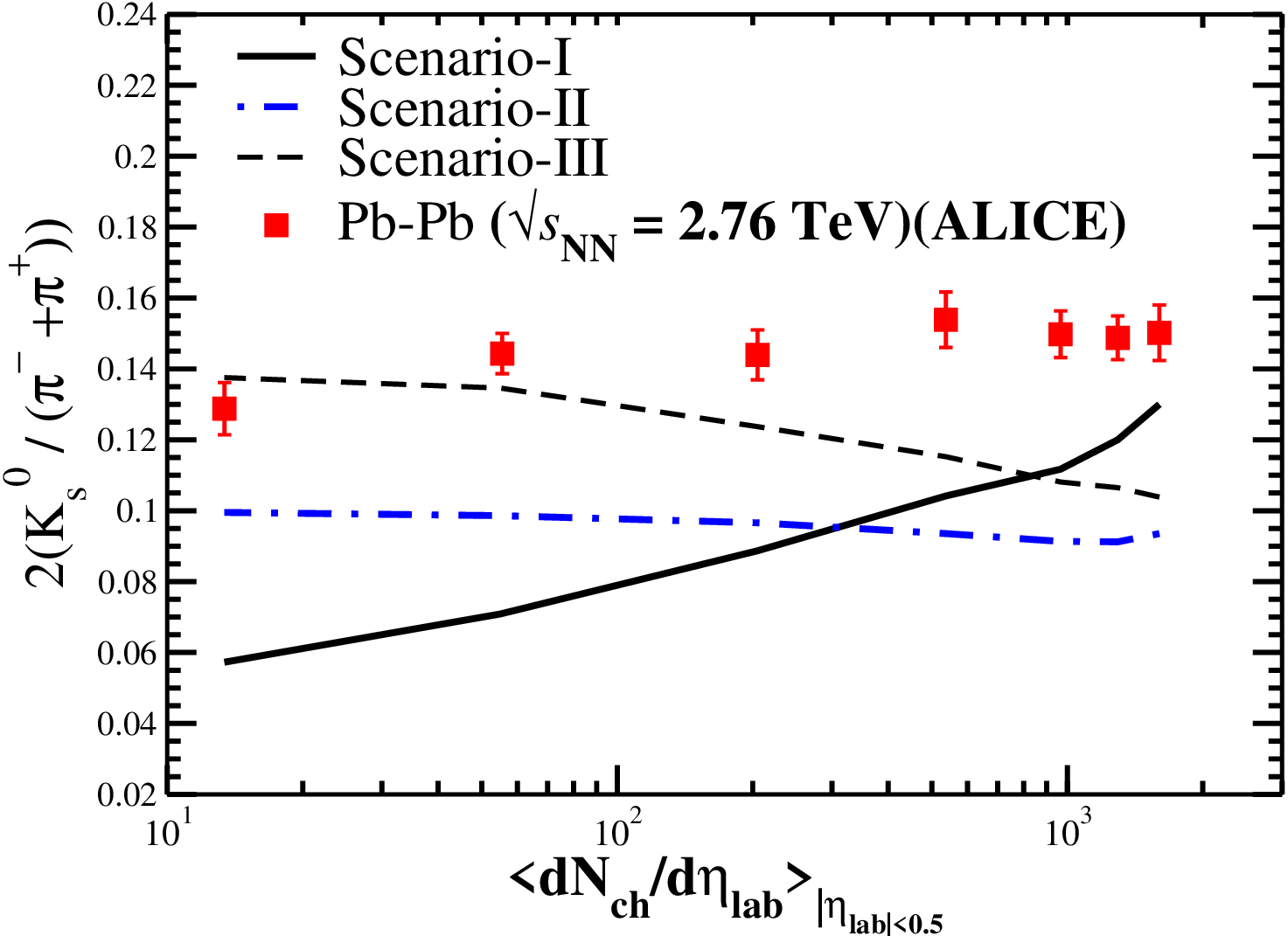}}\\
\subfloat{\includegraphics[width=0.42\textwidth,height=5cm]{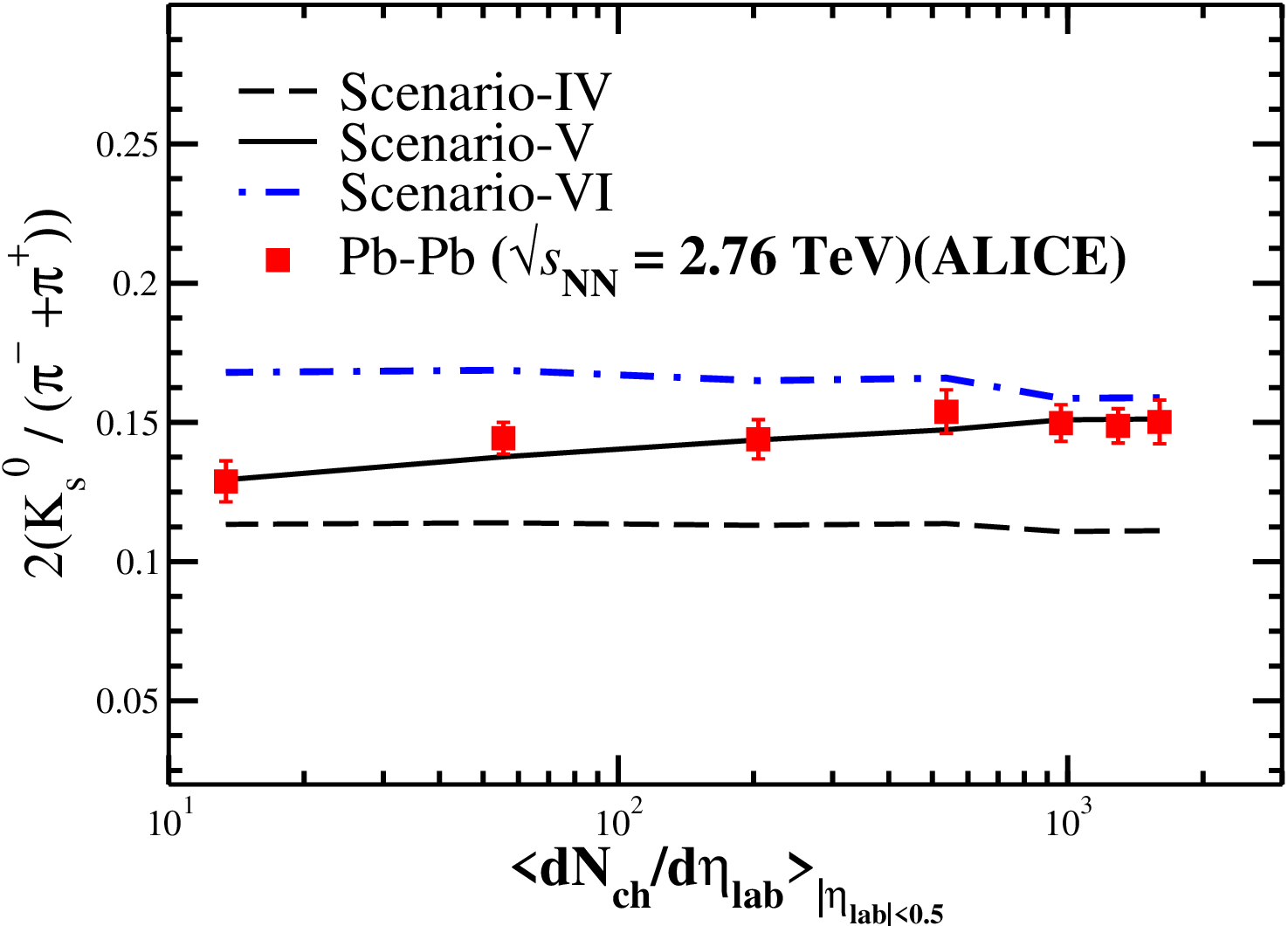}}
\caption{Yield ratio for $K_s^0$ from 2.76 TeV Pb+Pb collisions. The solid points with error bar are are the data points measured by ALICE collaboration. The solid/dashed/dotted lines are the results of theoretical calculation with different initial conditions for various scenarios. Left panel is for scenario-I, II, III and right panel is for scenario IV, V and VI}
\label{fig_singlestrangebypiratio-1}
\end{figure}

\begin{figure}
\centering
\subfloat{\includegraphics[width=0.42\textwidth,height=5cm]{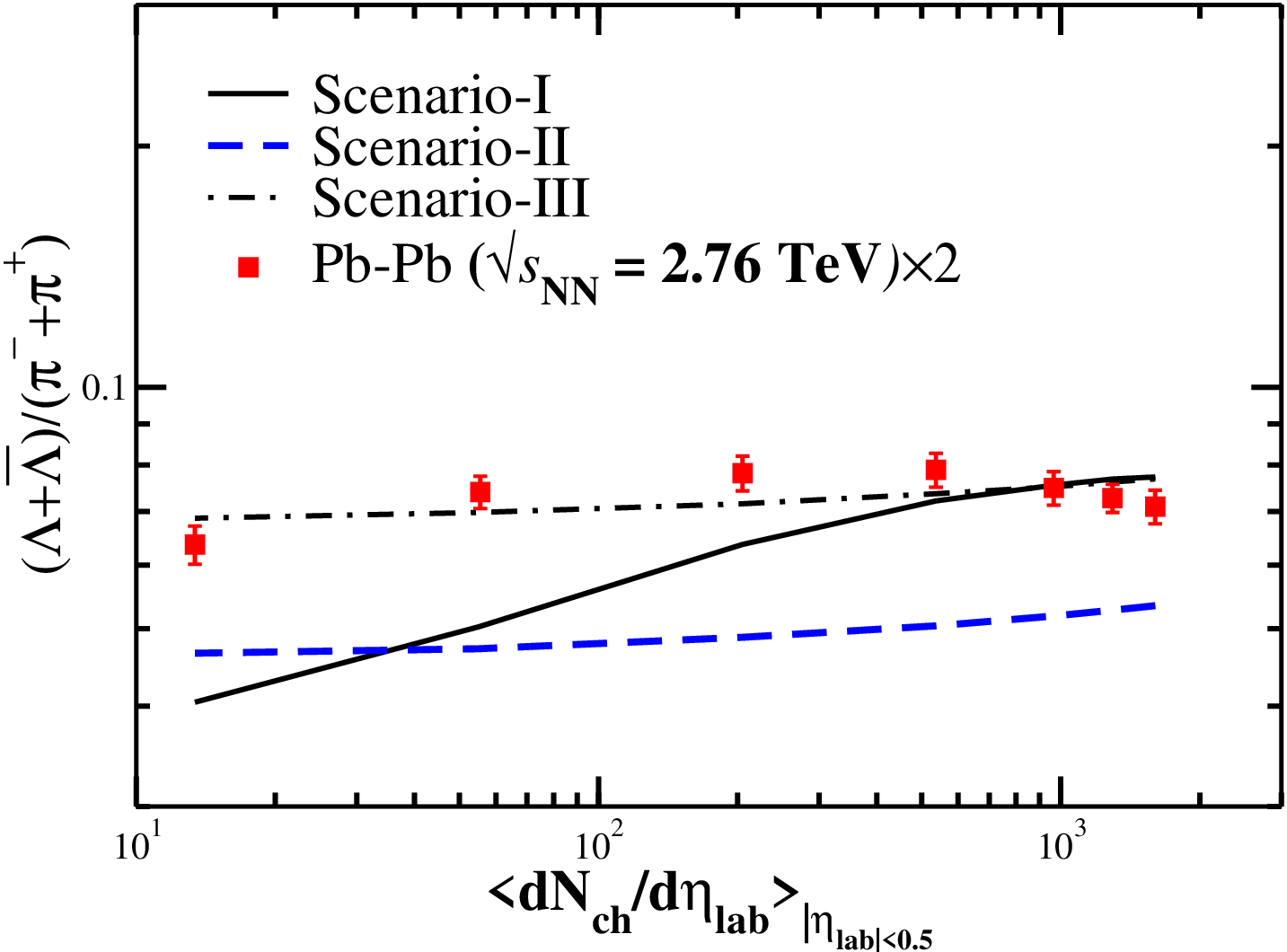}}\\
\subfloat{\includegraphics[width=0.42\textwidth,height=5cm]{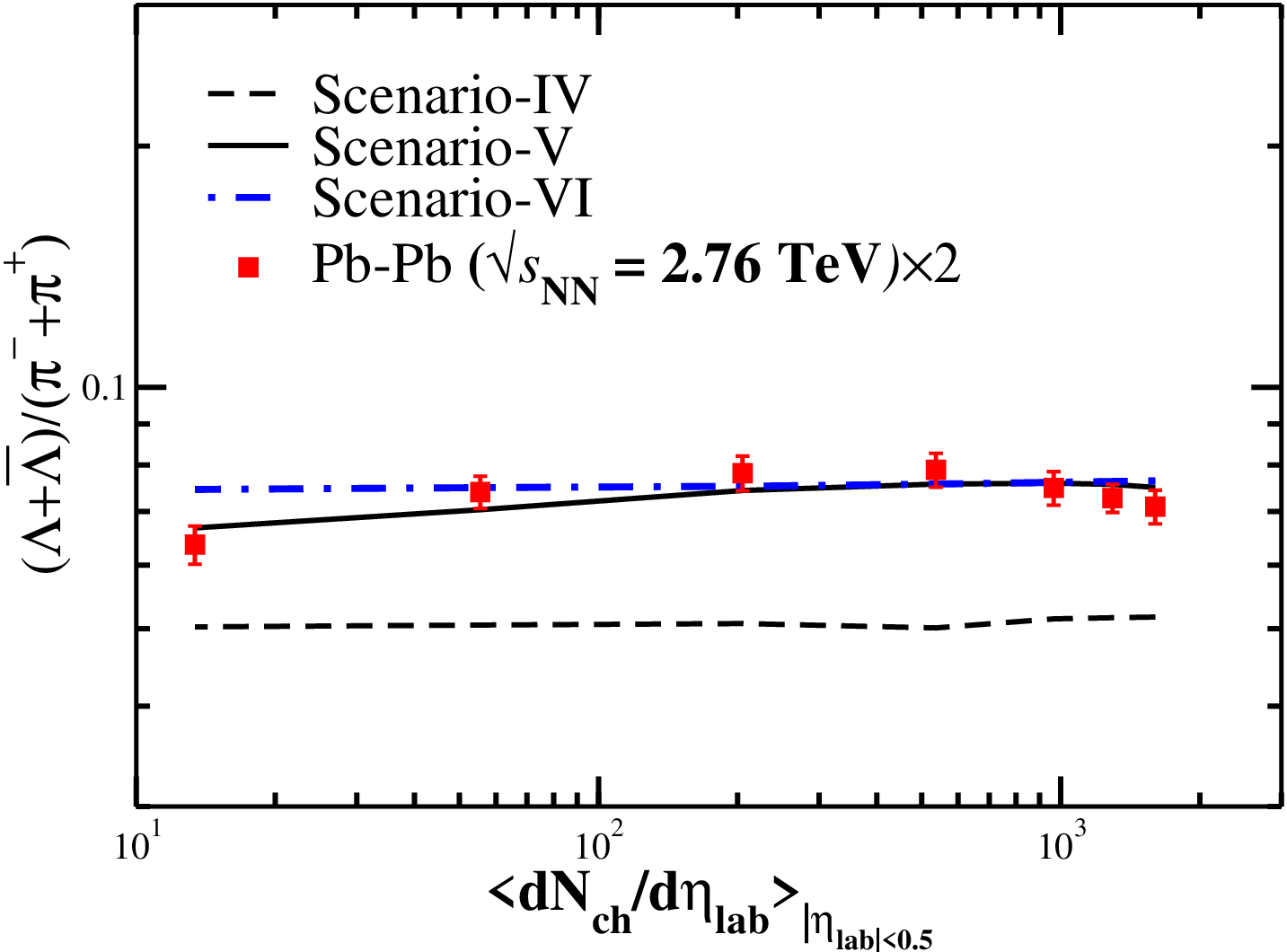}}
\caption{Yield ratio for $\Lambda$ from 2.76 TeV Pb+Pb collisions. The solid points with error bar are are the data points measured by ALICE collaboration. The solid/dashed/dotted lines are the results of theoretical calculation with different initial conditions for various scenarios. Left panel is for scenario-I, II, III and right panel is for scenario IV, V and VI}.
\label{fig_singlestrangebypiratio-2}
\end{figure}

In all scenarios we have taken $c_s^2=1/5$, $T_c=156 MeV$ and evaluated the yield ratio by stopping the calculation at $T_F$ and finally compared with the available data \cite{alicenature17,multistrange_alice_plb14}. Here we have considered $2K_s^0=K^+ + K^-$ as shown in Figs.\ref{fig_singlestrangebypiratio-1}, \ref{fig_singlestrangebypiratio-2}. The data of lightest hyperon, $\Lambda$ contains both $\Lambda^0$ and $\Sigma^0$. As it is difficult to separate $\Sigma^0$ from $\Lambda^0$ data. The isospin conservation channel of $\Sigma^0$ decay is $\Sigma^0\rightarrow\Lambda^0 +\pi^0$. But $\Sigma^0$ is not heavy enough to decay through this channel. Mass of $\Lambda^0$ and $\pi$ is more than $\Sigma^0$. Hence, $\Sigma^0$ preferably decays to $\Lambda^0$ and $\gamma$(branching ratio more than 99\%), which is an isospin non-conserving channel(weak decay) and difficult to reconstruct. Hence $\Lambda$ data contains $\Sigma^0$. Apart from that $\Lambda$ may contain the feed down from weak decays of $\Xi$ but it is already excluded from the data as mentioned in \cite{alice_abelev_prl12}. But the feed down contribution from $\Omega$ and other resonances such as $\Sigma^*(1385)$ family: $\Sigma^{+*}, \Sigma^{0*}, \Sigma^{-*}$ and $\Sigma^*(1660)$ are not removed from the data. In fact $\Sigma(1385)$ can decay to $\Lambda$ through isospin conservation channel. In our calculation, this contribution is taken care by multiplying a constant factor 0.8 to lambda production and adding it to the net yield.   
 
In scenario-I, $n_i$ for Kaon, Lambda, Sigma, Cascade and Omega are assumed to be 15\% away from the equilibrium value. The initial temperature $T_i$ for this scenario, in fact for all scenarios, is taken to be $T_c$ which is 156 MeV, a value taken from recent first principle calculation based on lattice computation($154\pm 9 MeV$). Velocity of sound is considered to be $c_s^2=1/5$, which is reasonable for hadron phase. The freeze out temperatures are different for different multiplicities as shown in Table-\ref{table_scenario_1234}. $T_F$ decreases with multiplicity in scenario-I and the results for {\it yield ratio } are shown in Figs.\ref{fig_singlestrangebypiratio-1} and \ref{fig_singlestrangebypiratio-2}. It does not explain all the data points for $K^0_s$ and $\Lambda$. This is similar to the {\it yield ratio} of $\Xi$ and $\Omega$ which are displayed in \cite{ghosharxiv19}.
In scenario-II, the system is allowed to evolve with an initial density 40\% away from the equilibrium value and with a constant $T_F$=144 MeV for all values. We have considered this freeze out temperature which is a lower temperature compared to the $T_F$ obtained from statistical hadronisation model(~154 MeV). We can't consider a higher temperature as it would exceed $T_c$. The $\tau_i$ is taken same for all scenarios for a particular multiplicity. $\tau_i$ is different for different multiplicity. Like previous scenario, Scenario-II also does not explain the kaon, lambda, cascade and omega data. Scenario-III under predicts the data(for all species).

\begin{figure}
\centering
\subfloat{\includegraphics[width=0.42\textwidth,height=5cm]{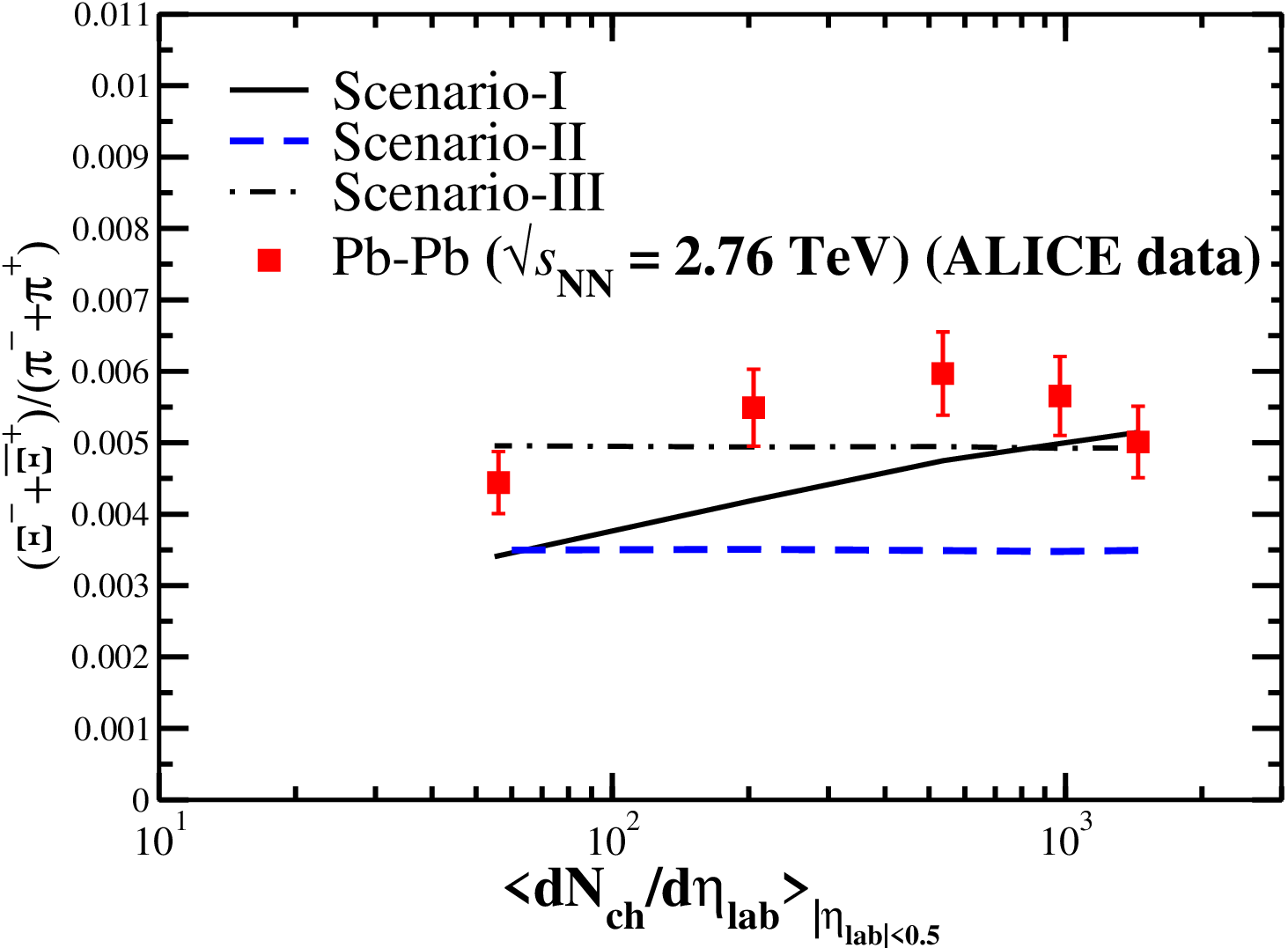}}\\
\subfloat{\includegraphics[width=0.42\textwidth,height=5cm]{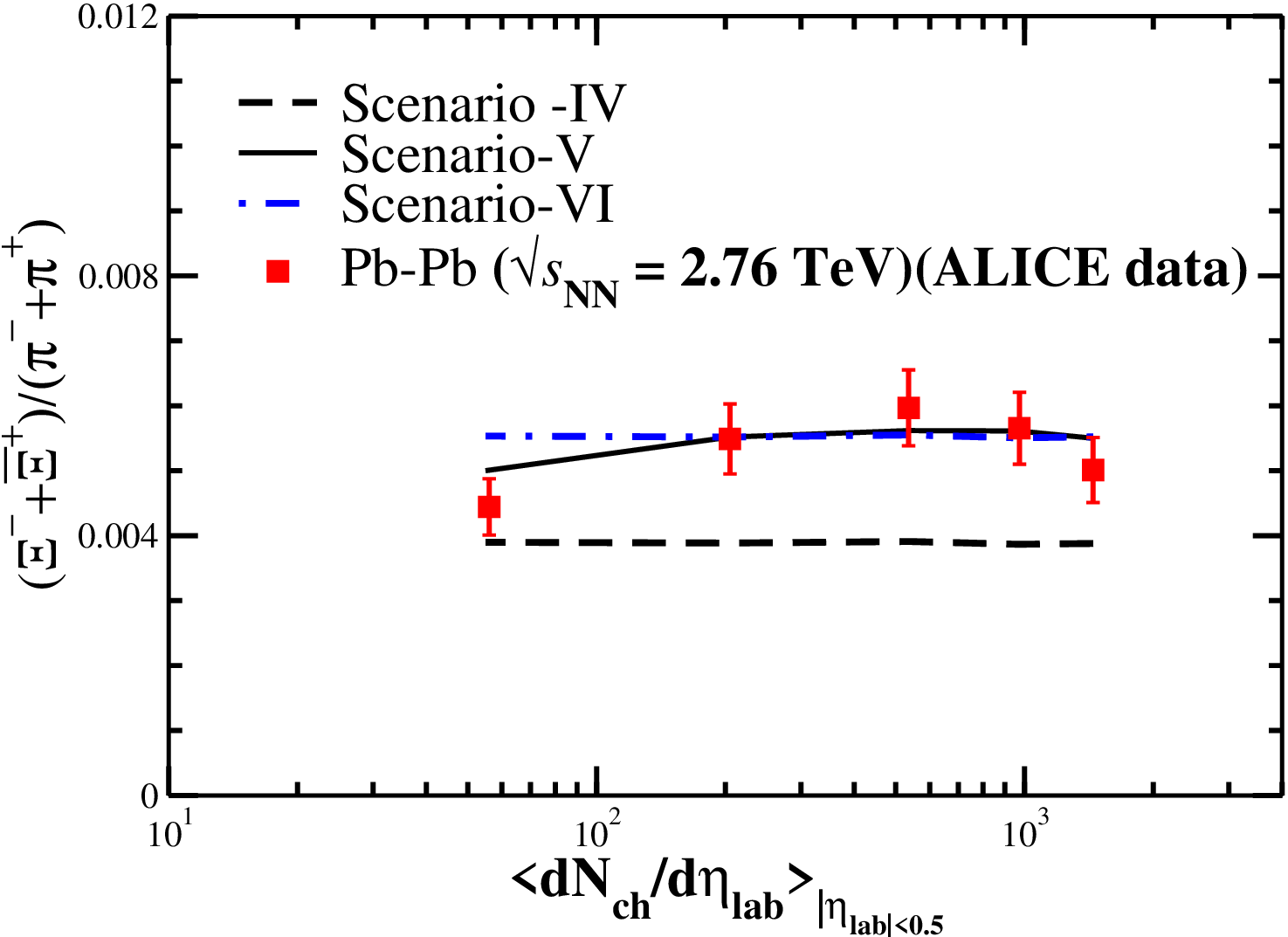}}
\caption{Yield ratio for $\Xi$ from 2.76 TeV Pb+Pb collisions. The solid points with error bar are are the data points measured by ALICE collaboration. The solid/dashed/dotted lines are the results of theoretical calculation with different initial conditions for various scenarios. Top panel is for scenario-I, II, III and Bottom panel is for scenario IV, V, VI}.
\label{fig_multistrangebypiratio-1}
\end{figure}

In scenario-III, we take $n_i$ to be 15\% away from equilibrium value with constant $T_F=144$ MeV, we observed data are under predicted (Figs.\ref{fig_singlestrangebypiratio-1} \& \ref{fig_multistrangebypiratio-1}.)

Being inspired for a $T_F$=154 MeV for all $dN_{ch}/d\eta$, as predicted by statistical hadronisation model for 2.76 TeV, LHC energy and as shown in the article by ALICE collaboration \cite{alice_adam2016}, we take $T_F$=154 MeV for all centralities with $n_i$=40\% away from equilibrium value ($n_i=0.6n_{eq}(T_i)$) in scenario IV and $n_i=15\% $ away from equilibrium value ($=0.85n_{eq}(T_i)$) in scenario-VI and tried to analyse the data. For scenario-IV all data of all species ($K, \Lambda, \Xi, \Omega$) are under predicted. But in case of scenario-VI, data of $\Lambda, \Xi, \Omega$ at higher multiplicities are explained (although not better). However, kaon data are over predicted. The initial conditions for scenario I-IV are tabulated in Table\ref{table_scenario_1234}.

\begin{figure}
\begin{center}
\includegraphics[scale=0.3]{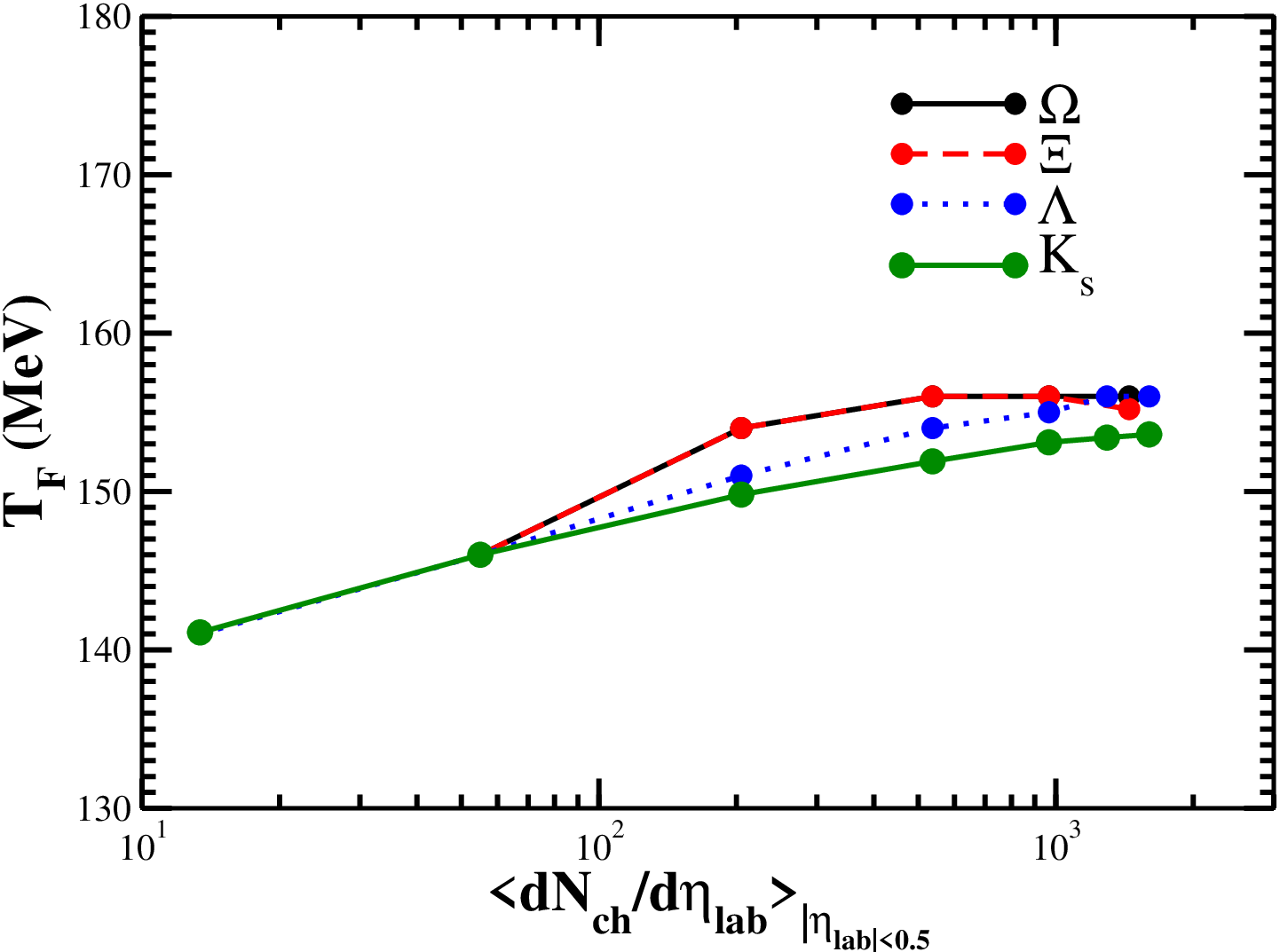}
\caption{Chemical freeze out temperatures for $K, \bar{K}, \Lambda, \Xi, \Omega$ extracted from scenario-V for various $dN_{ch}/d\eta$ considering {\bf Bjorken} expansion. }
\label{fig_tfvsdnchdeta}
\end{center}
\end{figure}

We tried to analyse for a scenario which could explain the data of all strange hadrons simultaneously and tried to get the information of $T_F$. That is scenario-V where $n_i$ is 15\% away from the equilibrium value. Here the freeze out temperatures that explain the {\it yield ratio} data for all $dN_{ch}/d_{\eta}$ are tabulated in Table\ref{table_scenario5}, which show a decreasing pattern of $T_F$ with multiplicity. These are displayed in Fig.\ref{fig_singlestrangebypiratio-2}. The simultaneous explanation of the {\it yield ratios} of multi-strange hadrons $\Xi$ and $\Omega$ for scenario-V is also displayed in Fig.\ref{fig_tfvsdnchdeta}. 
\begin{figure}
\centering
\subfloat{\includegraphics[width=0.42\textwidth,height=5cm]{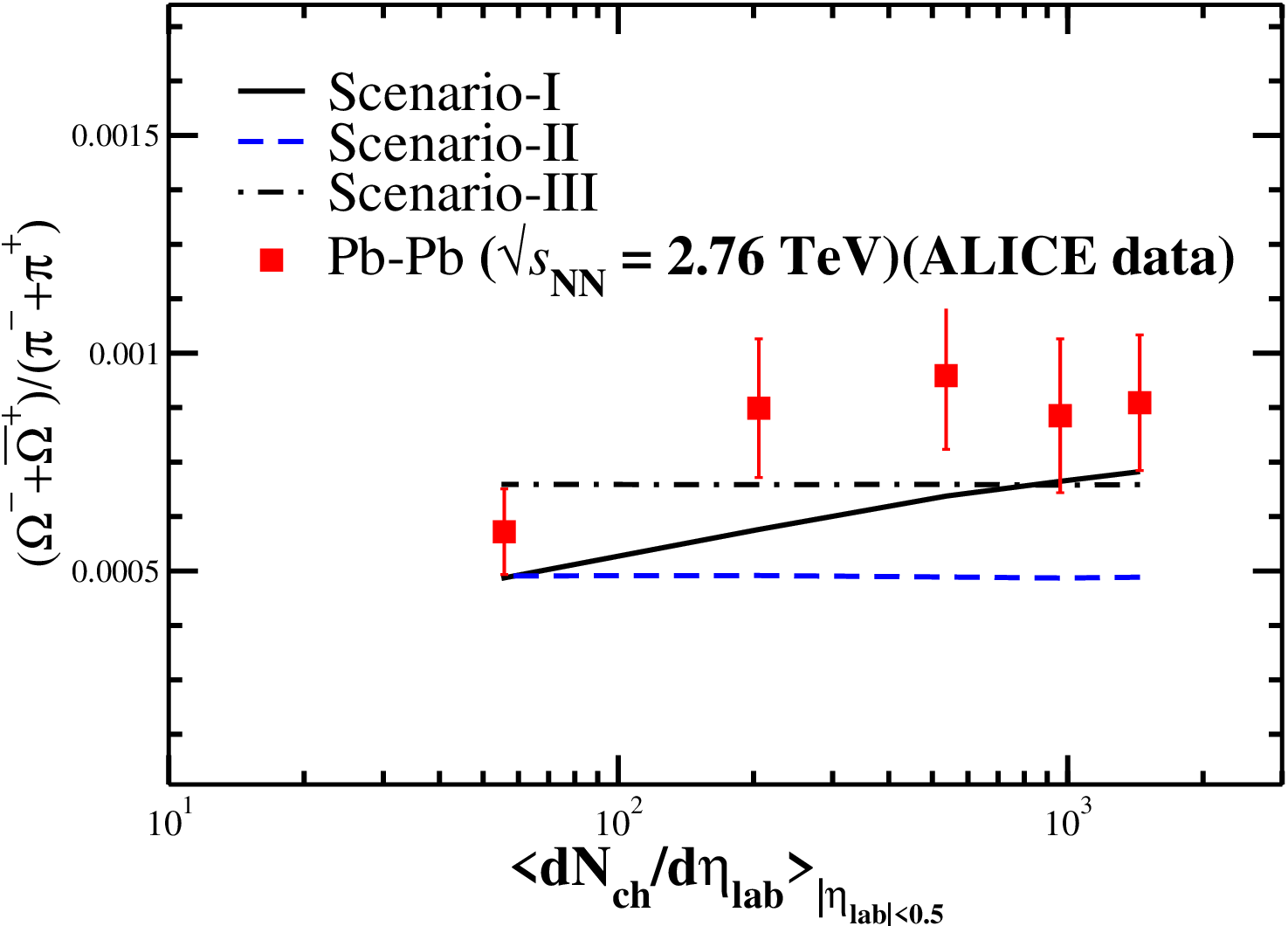}}\\
\subfloat{\includegraphics[width=0.42\textwidth,height=5cm]{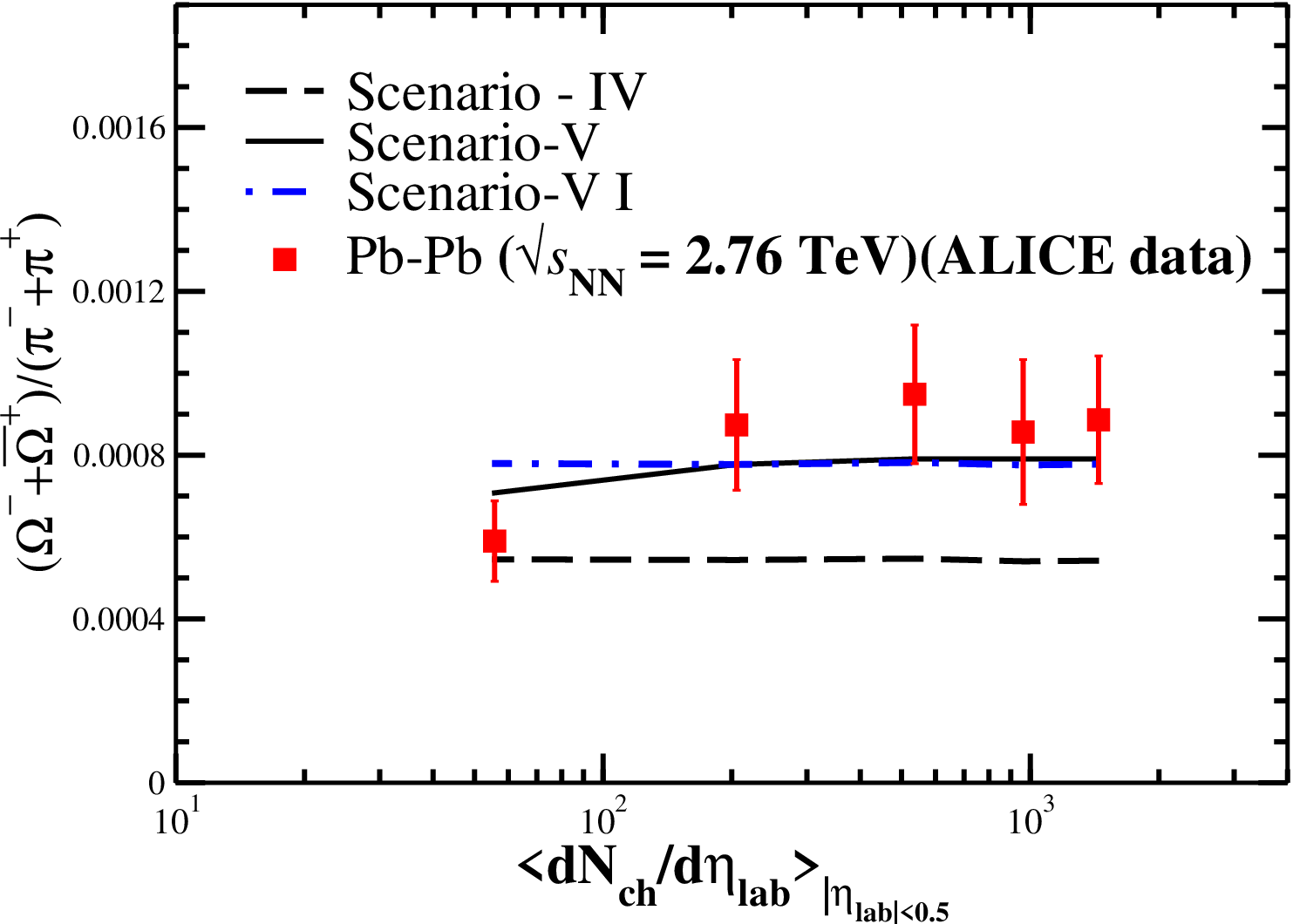}}
\caption{Yield ratio for $\Omega$ from 2.76 TeV Pb+Pb collisions. The solid points with error bar are are the data points measured by ALICE collaboration. The solid/dashed/dotted lines are the results of theoretical calculation with different initial conditions for various scenarios. Top panel is for scenario-I, II, III and Bottom panel is for scenario IV, V, VI}.
\label{fig_multistrangebypiratio-2}
\end{figure}
\section{\label{sec:summary} Summary}
The {\it yield ratio} of strange hadrons; $(K^++K^-)/(\pi^++\pi^-)$, $(\Lambda+\bar{\Lambda})/(\pi^++\pi^-)$, $(\Sigma+\bar{\Sigma})/(\pi^++\pi^-)$, $(\Xi^-+\bar{\Xi^+})/(\pi^++\pi^-)$ and $(\Omega+\bar{\Omega})/(\pi^++\pi^-)$ measured from p-p, p-Pb and Pb-Pb collisions at various centralities and colliding energies are presented by ALICE collaboration as an observable in \cite{alicenature17,multistrange_alice_plb14} against the charged particle multiplicity. The smooth rise of {\it yield ratio} pose a question- does the yield depend explicitly on multiplicity only? Does the colliding system, whether nucleon-nucleon (p-p)or nuclei(heavy)-nuclei(heavy) not matter? Do the colliding energies, $\sqrt{s_{NN}}$=2.76 TeV or 5.02 TeV or 7 TeV matter for the yield explicitly? Answering these questions in a single step is difficult. 

With an aim to answer these questions and to explain the strange hadron yields, we have made an initial framework and studied the strange hadron productions at LHC energy, $\sqrt{s_{NN}}$=2.76 TeV from Pb-Pb collisions microscopically, considering the cross sections of various interactions producing strange hadrons. We have calculated for LHC energy initially, because (i) measurements are available and (ii) the systems which are produced at various multiplicities of LHC energy have a common feature like negligible baryon chemical potential. The calculation would be extended to other colliding energies with different colliding systems.

In this article we have calculated the rate of single and multi-strange hadron productions considering various possible hadronic interactions and their cross sections, where most of the cross sections were constrained experimentally. Then the yield of $K, \bar{K}, \Lambda, \Xi, \Omega$ are evaluated solving rate equations simultaneously by considering the evolution of temperature and baryonic chemical potential of the system. Considering a hadronic system at $T_c$=154 MeV we have calculated the strange hadron yield with various initial conditions and obtained the yield ratio by normalising with thermal pions and finally compared the results with experimental observations to have an information of freeze out(chemical) scenario. The best explanation of the yield ratio data ({{\bf scenario-V}) at 2.76 TeV LHC energy suggests that (i)  multi-strange hadrons $\Xi, ~\Omega$ freeze out close to $T_c$, so also $\Lambda$ at higher multiplicity, (ii) The freeze out temperature of $K$ is different and less than multi strange hadrons, (iii) $T_F$ increases with multiplicity. This is for all strange hadrons.  At highest multiplicity, a single freeze out scenario for $K, \bar{K},\Lambda, \Xi, \Omega$ can be inferred. But the microscopic calculation suggests for the sequential freeze out of strange hadrons as the cross sections or rate of productions of hadronic species are different and follows an order. But the sequential freeze out is not clearly visible for strange hadron species at LHC energy. However it is expected at lower colliding energies. At LHC, probably the energy density and temperature is too high and the rate of production does not distinguish the differences in mean free paths of the species which lead to a common freeze out at high multiplicity.

A smooth change of $T_F$ with $dN_{ch}/d\eta$ at LHC energy is expected if the yield ratio depends only on $dN_{ch}/d\eta$ or ${\text N}_{part}$. Present calculation is expected to help to check whether energy density or finite size of the freeze out volume can be another parameter. Further improvement of calculation can be done by considering the corrections due to the volume of pion freeze out surface and considering the error bars due to the uncertainty of parameters. 

Thus it would be interesting to analyse the yield ratio data for all colliding energies available with a wide range of multiplicities to have a general conclusion in future. This microscopic work set a frame work to look for a better answer in future calculation. 

We have considered $c_s^2$=1/5 in our calculation and it explain the data nicely. When $c_s^2$=1/3 is considered the theoretical estimate overestimates the experimental observations for all $dN_{ch}/d\eta$. It would be more appropriate to use the parametrisation of the equation of state from lattice with temperature dependent $c_s^2 (T)$ which may improve the calculation. The yield of $\Xi,\Omega$ including single strange hadrons $K, \bar{K}, \Lambda$ are explained with this slow equation of state with $c_s^2$=1/5. 

Finally, summarizing the results it can be said that the strange hadrons $\Lambda, \Xi, \Omega$ freeze out earlier at a temperature close to $T_c$ at LHC energy, but Kaons freeze out little later because of its higher crosssection. But there is subtle and indistinguishable difference in freeze out temperatures of hyperons. Sequential freeze out or the differences in freeze out temperatures of various species may be clearly distinguishable or visible when the system is formed at lower colliding energy with substantial dominance from baryons. If we move from central collision to peripheral collisions, the freeze out temperature does not depend strongly on centrality. It is same for $\Lambda, \Xi$ and $\Omega$. For kaon it differs slightly. In general, for particular species, The freeze out temperature does not depend strongly on centrality. 
\par
{\bf Acknowledgment:} 
Author P. Ghosh thanks Dr. J. Alam for the financial support from CNT project vide no. 3/5/2012/VECC/R\&D-I/14802 during the stay at VECC.
\appendix 
\section{Rate Equation}
We first outline the derivation of the chemical rate equation for the evolution of number density of particle type $a$ ~\cite{kolb}. The Boltzmann equation is given by
 \begin{equation}
  p^\mu \partial _\mu f_{a} = C[f_{K}]
 \end{equation}
 where $f_{a}(x,p,t)$ is the phase space density of species $a$. Assuming the phase space density to be spatially homogeneous and 
 isotropic we have
 \begin{equation}
  E\frac{\partial f_{a}}{\partial t} = C[f_{K}]
 \end{equation} 
 Integrating over momenta we get
 $$\frac{g_a}{(2\pi)^3} \int d^3 p\frac{\partial f_{a}}{\partial t} = \frac{g_a}{(2\pi)^3} \int \frac{d^3p}{E} \, C[f_{a}]$$
 which gives
 \begin{equation}
 \label{eqn::chem_rate_eqn_step1}
  \frac{dn_a}{dt}=\frac{g_a}{(2\pi)^3} \int \frac{d^3p}{E} \, C[f_{a}]
 \end{equation}
 where $n_a (t)$ is given by
 $$n_a(t) = \frac{g_a}{(2\pi)^3} \int d^3 p \, f_{a}(E,t)$$
 Let us define
 $$d\Pi =\frac{g}{(2\pi)^3} \frac{d^3p}{E} $$
 so that the RHS of Eq.(\ref{eqn::chem_rate_eqn_step1}), for some reaction $a+b\rightarrow c+d$, may be written as (assuming classical particles)\footnotesize
 \begin{align*}
  \int d\Pi_a C[f_a] = -\int &d\Pi_a \, d\Pi_b \, d\Pi_c \, d\Pi_d \, (2\pi)^4 \delta ^4(p_a+p_b-p_c-p_d) F_{abcd}
 \end{align*}\normalsize
 where $F_{abcd}$ is given by
 $$F_{abcd} = |\mathcal{M}|^2_{a+b\rightarrow c+d}f_{a}f_b-|\mathcal{M}|^2_{c+d\rightarrow a+b}f_cf_d$$
 and $\mathcal{M}_{a+b\rightarrow c+d}$ denotes the amplitude for forward reaction $a+b\rightarrow c+d$ and $\mathcal{M}_{c+d\rightarrow a+b}$ denotes the amplitude for the reverse reaction $c+d\rightarrow a+b$. Assuming $PT$ invariance, we have
 $$|\mathcal{M}|^2_{a+b\rightarrow c+d}=|\mathcal{M}|^2_{c+d\rightarrow a+b}=|\mathcal{M}|^2$$
 so that we get \footnotesize 
 \begin{align*}
  \int d\Pi_a C[f_a] &= -\int d\Pi_a \, d\Pi_b \, d\Pi_c \, d\Pi_d \, (2\pi)^4 \\
  & \times \delta ^4(p_a+p_b-p_c-p_d) |\mathcal{M}|^2  (f_{a}f_b-f_cf_d)
 \end{align*} \normalsize
 The differential cross-section~\cite{peskin_book} for the reaction $a+b\rightarrow c+d$ is given by
 $$d\sigma = \frac{1}{E _a\, E_b\, v_{ab}}\int d\Pi_c \, d\Pi_d \, (2\pi)^4\delta ^4(p_a+p_b-p_c-p_d) \, |\mathcal{M}|^2$$
 where $v_{ab} = |v_a-v_b|$ denotes the Moller velocity (or relative velocity in loose terms) and which is given by
 $$v_{ab} = \frac{\sqrt{(p_a.p_b)^2-m_a^2m_b^2}}{E_aE_b}$$
 The above expressions suggest the definition for the non-thermal (NTh) averaged cross section times velocity as
 \begin{align}
  \langle \sigma_{ab}\, v_{ab}\rangle _{NTh} &= \frac{1}{n _a\, n_b}\int d\Pi_a \, d\Pi_b \, d\Pi_c \, d\Pi_d \, (2\pi)^4 \nonumber \\
  & \times\delta ^4(p_a+p_b-p_c-p_d) \, |\mathcal{M}|^2 f _a \, f_b\label{eqn_reaction_rate_K_Nth}
 \end{align}
 Hence the evolution equation for number density of particles $a$ will be 
 \begin{equation}
  \frac{dn_a}{dt} = -n_a n_b \, \langle \sigma_{ab} \, v_{ab}\rangle _{NTh}  + n_c n_d \, \langle \sigma _{cd}\, v_{cd}\rangle _{NTh}
 \end{equation}
 To make further progress we assume that the non-thermal reaction rate is approximately equal to the thermal average \emph{i.e.} near chemical equilibrium and the same is also assumed for slightly away from equilibrium. $\langle \sigma \, v\rangle _{NTh} \approx \langle \sigma \, v\rangle _{Th}$. Hence the chemical rate equation becomes
 \begin{equation}
  \frac{dn_a}{dt} = -n_a n_b \, \langle \sigma_{ab} \, v_{ab}\rangle _{Th}  + n_c n_d \, \langle \sigma _{cd}\, v_{cd}\rangle _{Th}
 \end{equation} 
 From now onwards we will remove the subscripts ``Th'' for thermal averages. All averages that appear below should be understood as thermal averages. If the system is also expanding, then the rate equation becomes
 \begin{equation}
  \frac{dn_a}{dt}+\Gamma_{exp} \, n_a = -n_a n_b \, \langle \sigma_{ab} \, v_{ab}\rangle _{Th}  + n_c n_d \, \langle \sigma _{cd}\, v_{cd}\rangle _{Th}
 \end{equation}  
 where $\Gamma_{exp}$ is the expansion rate. For (1+1)-dimensional Bjorken expansion, $\Gamma_{exp}=\frac{1}{t}$.  Also, since we are studying the production of strange hadrons whose masses are much larger than the temperature range $(T<T_c)$ in which we are interested, we can take the equilibrium distribution to be Maxwell-Boltzmann distribution. Hence the reaction  rate, $R(T)$, for a reaction with incoming particles $a,b$ and outgoing particles $c,d$ and at a temperature $T$ is given by ~\cite{kapusta86,gondolo91},
 \begin{align}
  \langle \sigma v\rangle_{ab\rightarrow cd} &= \frac{\int \, \sigma v \, e^{-E_a/T}e^{-E_b/T} \, d^3{\bf p_a} \, d^3{\bf p_b}}{\int \, e^{-E_a/T}e^{-E_b/T} \, d^3{\bf p_a} \, d^3{\bf p_b}} \nonumber \\
  &=\frac{\mathcal{C}_{ab}(T)}{16\pi^2T^2}\int \, \sigma v \, e^{-E_a/T}e^{-E_b/T} \, d^3{\bf p_a} \, d^3{\bf p_b}
  \label{eqn_reactionrate_app}
 \end{align}
 where $\mathcal{C}_{ab}(T)$ is given by the following expression 
 $$\mathcal{C}_{ab}(T) = \frac{1}{m_a^2m_b^2K_2(m_a/T)K_2(m_b/T)},$$
 and $K_2(x)$ denotes the modified bessel function of second kind. Taking a preferential direction for $\textbf{p}_a$ along $z$-direction and taking $\theta$ as the angle between $\textbf{p}_a$ and $\textbf{p}_b$, one gets after simplification \footnotesize
 \begin{align}
  \langle \sigma v\rangle_{ab\rightarrow cd} &= \frac{\mathcal{C}_{ab}(T)}{8T}\int _{s_0}^\infty ds  [s-(m_a+m_b)^2][s-(m_a-m_b)^2] \nonumber \\
  & \hspace{0.2in}\times \frac{1}{\sqrt{s}} \sigma \, K_1(\sqrt{s}/T)
 \end{align}\normalsize
 where $K_1(\sqrt{s}/T)=\frac{\sqrt{s}}{T}\int \, dE_+\, e^{-E_+/T}\sqrt{E_+^2-s}$. With $z=\frac{\sqrt{s}}{T}$, one can write the thermal averaged reaction rate as
 \footnotesize
 \begin{align}
  \langle \sigma v\rangle_{ab\rightarrow cd} &=\frac{T^4}{4}\mathcal{C}_{ab}(T)\int _{z_0}^{\infty} \, dz\, [z^2-(m_a/T+m_b/T)^2]\nonumber\\
  & \hspace{0.2in}\times [z^2-(m_a/T-m_b/T)^2]\sigma K_1(z)
  \label{eqn_reacrate_app}
 \end{align}\normalsize
 where $z_0=\text{max}(m_a+m_b,m_c+m_d)/T$.

\end{document}